\documentclass[10pt,twocolumn]{article}
\usepackage{graphicx} % Required for inserting images

\usepackage{authblk}
\usepackage{float}
\usepackage{braket}
\usepackage{mathtools}
\usepackage{array}
\usepackage[acronym]{glossaries}
\usepackage[table,xcdraw]{xcolor}

\usepackage{amsmath}
\usepackage{mathrsfs}
\usepackage{xcolor}
\usepackage{amssymb}
\usepackage{mathabx}

\usepackage{url}
\usepackage{makecell}
\usepackage{latexsym}
\usepackage{physics}
\usepackage{tikz}
\usepackage{comment}
\usepackage[font=small,labelfont=bf]{caption}
\usepackage{subcaption}
\usepackage[hidelinks]{hyperref}
\usepackage[noend]{algpseudocode}
\usepackage{algorithm}
\usepackage{algpseudocode}
\usepackage{bbold}
\usepackage{changepage}
\usepackage[normalem]{ulem}
\usepackage[titletoc,toc,title]{appendix}

\usetikzlibrary{quantikz2}

\usepackage[sorting = none, natbib=true,style=numeric-comp, maxbibnames=20]{biblatex}
\newcommand{\braketmatrix}[3]{\left \langle #1 \middle| #2 \middle| #3 \right \rangle}
\algrenewcommand\algorithmicrequire{\textbf{Input:}}
\algrenewcommand\algorithmicensure{\textbf{Output:}}

\title{Multi-QIDA method for VQE state preparation in molecular systems}

\author[1,2]{Fabio Tarocco}
\author[1,2] {Davide Materia}
\author[1,2] {Leonardo Ratini}
\author[2, $\dagger$]{Leonardo Guidoni}

\affil[1]{Dipartimento di Ingegneria e Scienze dell'Informazione e Matematica\\ Universit\`a degli Studi dell'Aquila, Coppito, L'Aquila, Italy}

\affil[2]{%
Dipartimento di Scienze Fisiche e Chimiche, Universit\`a degli Studi dell’Aquila, Coppito, L’Aquila, Italy }%

\affil[$\dagger$]{Email: leonardo.guidoni@univaq.it}

\date{\today}

\begin{document}

\twocolumn[
\begin{@twocolumnfalse}
\maketitle
\begin{abstract}

The development of quantum algorithms and their application to quantum chemistry has introduced new opportunities for solving complex molecular problems that are computationally infeasible for classical methods. In quantum chemistry, the Variational Quantum Eigensolver (VQE) is a hybrid quantum-classical algorithm designed to estimate ground-state energies of molecular systems. Despite its promise, VQE faces challenges such as scalability issues, high circuit depths, and barren plateaus that make the optimization of the variational wavefunction. 
To mitigate these challenges, the Quantum Information Driven Ansatz (QIDA) leverages Quantum Mutual Information (QMI) to construct compact, correlation-driven circuits. In this work, we go back to the original field of application of QIDA, by applying the already defined Multi-Threshold Quantum Information Driven Ansatz (Multi-QIDA) methodology on Molecular Systems. to systematically construct shallow, layered quantum circuits starting from approximate QMI matrices obtained by Quantum Chemistry calculations.
The Multi-QIDA approach combines efficient creation of the QMI map, reduction of the number of correlators required by exploiting Minimum/Maximum spanning tress, and an iterative layer-wise VQE optimization routine. These enhancements allow the method to recover missing correlations in molecular systems while maintaining computational efficiency. Additionally, the approach incorporates alternative gate constructions, such as SO(4) correlators, to enhance the circuit expressibility without significantly increasing the circuit complexity.

We benchmark Multi-QIDA on systems ranging from small molecules like H$_2$O, BeH$_2$, and NH$_3$ in Iterative Natural Orbitals (INOs) basis set, to active-space models such as H$_2$O-6-31G-CAS(4,4) and N$_2$-cc-pVTZ-CAS(6,6), comparing it to traditional hardware-efficient ans\"atze. The results show that Multi-QIDA consistently outperforms the ladder topology ans"atze in terms of energy accuracy, correlation recovery, and resource utilization. 
Interestingly, in addition to the higher fidelity to the exact ground state, we observe a significant improvement of the quality of the variational wavefunction which better preserves the correct symmetries, such as $\hat{S_z}, \hat{S^2}$, and $\hat{N}$.
\end{abstract}
\end{@twocolumnfalse}
]
\clearpage

\section{Introduction}

The Variational Quantum Eigensolver (VQE) method utilizes a hybrid quantum-classical approach to estimate ground state energies of molecular systems by optimizing a parameterized quantum ansatz through iterative energy minimization. Since about a decade the method has been largely studied (and criticized) for its employnment into near-term applications designed for quantum chemistry~\cite{Peruzzo2014, McClean_2016, Kandala_2017, Cerezo_2021, fedorov2021vqe, Bharti_2022, Tilly_2022}.

Its simplicity instead hides scalability challenges because of to the growth of parameter space, the flatness of the corresponding energy landscape, the intensive exchange of data between quantum and classical devices, and circuit depth on noisy quantum devices which need to be mitigated. To address some of these issues, various improved algorithms have been proposed~\cite{Barkoutsos_2018, Grimsley_2019,Yordanov_2021,benfenati2021improved}. 

In this respect, a fundamental role is led by the choice of the parametrized wavefunction that is going to be used as a trial wavefunction together with the VQE procedure. The shape and the structure of the ansatz are non-trivial and system-dependent, and in general, two types of ans\"atze can be defined. The first class exploits wavefunctions directly constructed to leverage the characteristics of quantum hardware. This empirical approach, known as the Heuristic Ansatz~\cite{Kandala_2017, Ganzhorn_2019,rattew2020domainagnostic, Tkachenko_2021, Tang_2021}, comprises repetitions of blocks of parametrized rotations and entanglement gates. It is designed without relying on information about the physical system, focusing solely on exploiting the quantum hardware's capabilities. While the Heuristic Ansatz better utilizes the quantum hardware, it comes at the cost of losing the physical meaning associated with the wavefunction interpretation, with the advantage of considering significantly shallower circuits, providing a potential avenue to address scalability concerns. 
In contrast, the second family of ans\"atze involves translating classical Quantum Chemistry methods into the language of quantum computation. ~\cite{Hoffmann_1988,Cooper_2010,Evangelista_2011,romero2018strategies, Magoulas2023}.
One of the most important chemically inspired ansatz is the Unitary Coupled Cluster (UCC) which directly translated the wavefunction structure as defined for the "classical" Coupled cluster theory.~\cite{Hoffmann_1988,Cooper_2010,Evangelista_2011,romero2018strategies, Magoulas2023}

Regardless of the approach used to construct the ans\"atze, increasing complexity in parameterized quantum circuits (PQC) results in longer and deeper circuits, which not only heightens the risk of error accumulation but also contributes to the emergence of Barren Plateaus~\cite{larocca2024review}, an issue characterized by an exponentially flat optimization landscape. While adjustments in optimization techniques or error mitigation strategies do not directly resolve barren plateaus, employing shallow and adaptively structured ans\"atze has proven to be an effective countermeasure. Examples of such approaches include ADAPT-VQE, a variant of the UCC method, which builds a wavefunction only by exploiting energetically meaningful single and double excitations that are applied on the circuit in an adaptive way ~\cite{Grimsley_2019, Yordanov_2021, Tang_2021}. 
 Another approach is to optimize both the Hamiltonian and the wavefunction with the objective of compacting the circuit and reducing the depth, done by applying the WAHTOR algorithm~\cite{ratini_2022}.  
 Shallow circuits that can also encode chemically relevant information and being build, for example, following the Mutual Information present in the system, as done with Quantum Information Driven Ansatz~\cite{materia2023quantum}. This approach allows to define quickly initial guess with a limited number of CNOTs.
Obtaining energetically correct results with a good overlap with the ground state is not only useful as a starting guess for more complex ansatz construction protocol or other VQE-based procedures, but also as a trial wavefunction from which samples can be extracted. Recently, different works in which a non-VQE-based, mainly exploiting the quantum version of a Selected Configuration Interaction (SCI) approach, called Quantum SCI (QSCI)~\cite{kanno23}, Quantum Subspace Diagonalization (QSD) ~\cite{qsd22}, or Sample-Based Quantum Diagonalization (SQD)~\cite{quantumcentric24}, have been proving relevant industrial results and application of Quantum Algorithm to quantum utility-scale ~\cite{iron25,quantumcentric24}. In these works, the principal actor is the trial wavefunction from which the samples are drawn, which, for example, can be either a fixed Quantum Number Preserving (QNP) circuit ~\cite{qnp2} or a chemically inspired hardware-efficient Local Unitary Coupled Jastrow (LUCJ)~\cite{quantumcentric24}. Despite these advantages, being able to construct compact wavefunctions that can build effective trial wavefunctions that can be used to sample relevant determinants with the correct quantum symmetries is still a challenge.

With this work, we wanted to test the performance of the Multi-QIDA approach introduced by us for lattice spin-systems ~\cite{tarocco2024} on molecular systems. In particular, we combined the idea behind the usage of Natural Orbitals to obtain a more correlated and compact wavefunction and the application of multiple layers of QIDA-based circuits. We also introduced two different ways of reducing the number of correlators to insert in each QIDA-layer, namely, \textit{maximal correlation} and \textit{topological distance}, based on Spanning Trees. The approach has been integrated with the QMI builder tool SparQ ~\cite{materia2024} to efficiently compute reference Quantum Mutual Information matrices. 
With this new method, we have been able to partially recover the missing electron correlations by building and optimizing a layered-structured ans\"atz which is extended at each iteration by including a newly computed QIDA-layer based on the QMI. We have thus amplified the results obtained by the stand-alone QIDA ansatz in its same field of application by recovering mid/low-correlated qubit-pairs. In this way, we have been able to obtain shallow circuits that, compared to the same CNOT-count of a HEA ladder-fashion circuit, reach higher correlation energy with increased precision. Anticipating our results, the optimised ans\"atze turned out are also to be qualitatively better than simpler ones in terms of preserving the relevant spin and electronic symmetries and quantities of the exact ground states, like $\hat S_z$, $\hat S^2$, and $\hat N_e$.

This paper is split into the following sections. Firstly, with Section \ref{sec:background}, we provide a brief introduction on the theoretical background required for the full explanation of the Multi-QIDA approach. In Section \ref{ssec:vqe}, we introduce the Variational Quantum Eigensolver (VQE) in the context of Quantum Computing formalism. The main physical quantity used to build our ans\"atze Quantum mutual-information measure is defined in Section  \ref{ssec:qmi}. Then, the formalism from which we defined one of the basis sets used in this work, the Iterative Natural Orbitals, is defined in Section \ref{ssec:inos}. The extrapolation of the QMI matrices with SparQ is presented in Section \ref{ssec:sparq}. The full framework of Molecular-System Multi-QIDA is shown in Section \ref{sec:msmqida}. In particular,  the entangling gates used in this work, the \textit{SO4} gates, in Section \ref{ssec:so4}, while the selection criteria based on spanning tree to place them follows in Section \ref{ssec:selection_crit}. Thanks to the selected pairs, the ansatz is then built following the procedure explained in Section \ref{ssec:lb} and optimized accordingly to the optimization scheme of Section \ref{ssec:iterVQE}.  We then presented the system that have been studied in this work, both simple molecule and active-space selected systems, are presented in Section \ref{ssec:systems}, followed by the metrics used to evaluate the efficiency of the ans\"atze and the symmetry measures computed, in Section \ref{ssec:mm}. The simulation details are explained in Section \ref{ssec:sim_det}. Finally, in Section \ref{sec:res}, results are shown and explained.
\section{Background}
\label{sec:background}
In this section, we collected very briefly all the methods and related principles that have been used in our work. 
\subsection{Eigenproblem for Quantum Chemistry Hamiltonian}
\label{ssec:vqe}
Given a molecule and a set of spin molecular orbitals (MOs), the second-quantized electronic Hamiltonian is defined as follows \begin{equation}
\label{eq:H_elec}
\hat{H} = \sum_{ij} h_{ij} \hat{a}_i^\dagger \hat{a}_j^{} + \frac{1}{2} \sum_{ijkl} \Gamma_{ijkl}\hat{a}_i^\dagger \hat{a}_j^\dagger \hat{a}_k^{} \hat{a}_l^{},
\end{equation} where $a^{\dagger}_{i}$ and $a^{}_{i}$ are creation and annihilation operators, respectively, while $h_{ij}$ and $\Gamma_{ijkl}$ are the one-body and two-body integrals. In the summation, the indices of Equation \ref{eq:H_elec} iterate over the set of spin MOs that can be either composed by Hartee-Fock orbitals, or other types of orbitals such as Localized Orbitals (e.g. Boys~\cite{boys1960}, Pipek-Meyez~\cite{PipekMeyez1989},Edmiston-Ruedenberg~\cite{Edmiston1963}, etc.) or Natural Orbitals~\cite{Lowdin1956}.
By exploiting the Jordan-Wigner mapping~\cite{Jordan1928}, each fermionic operator can be mapped from the fermionic space to the space of the qubit, thus as a combination of quantum-gate operations. In particular, 
\begin{equation}
a^{\dagger}_{i} = Q_i^{\dagger}\prod^{i-1}_{j=0}{Z_j}
\label{eqn:a_dagger}\end{equation} and
\begin{equation}
a^{}_{i} = Q^{}_i\prod^{i-1}_{j=0}{Z_j},
\label{eqn:a}\end{equation} where $Q^{\dagger}_{i} = \frac{1}{2}(X_i - iY_i)$, $Q_{i} = \frac{1}{2}(X_i + iY_i)$, and $i$ iterates over the set of MOs. They can be associated to creation and annihilation operators in the qubit space, respectively. As well as the fermionic creation and annihilation operators act to modify the occupancy of the relative spin-MO, the qubit operators act to change the state of the qubit related to the specific spin-MO. The full correspondence is finalized with a series of Pauli-Z matrices that allow to compute the parity of the state and account for the fermionic anticommutation between $a$ and $a^{\dagger}$. Using Equation \ref{eqn:a_dagger} and Equation\ref{eqn:a_dagger} in Equation\ref{eq:H_elec}, the new Hamiltonian defined in the qubit space, $\hat{\mathbf{H}}$, can be written as 
\begin{equation}
\hat{\mathbf{H}} = \sum_i^L c_i \hat{P}_i = \sum_{i}^L c_i \bigotimes_{j=0}^{N_{MO}}{\hat{\sigma}_j^i},
    \label{eqn:H_elec_qubit}
\end{equation} where each $\hat{\sigma}_j \in \{X,Y,Z,I\}$ is a Pauli matrix acting on the $j$-th qubit, $c_i$ is a coefficient relative to the $i$-th Pauli string $\hat{P}_i$, $N_{MO}$ is the number of qubit or number of spin-MOs, and $L$, proportional to $N_{MO}^4$, is the number of Pauli string composing the qubit Hamiltonian.

Now, the Variational Quantum Eigensolver(VQE)~\cite{Peruzzo2014} can be used to solve the electronic structure problem. VQE is an hybrid quantum-classical algorithm that can be used to minimize the expectation value \begin{equation}
E(\bar{\theta}) = \langle \psi(\bar{\theta})|\hat{\mathbf{H}}|\psi(\bar{\theta})\rangle,
\label{eqn:cost_vqe}
\end{equation} which relies on the Rayleigh-Ritz~\cite{Ritz1909} variational principle \begin{equation}
\langle \psi(\bar{\theta})|\hat{\mathbf{H}}|\psi(\bar{\theta})\rangle \geq E_0
    \label{eqn:var_princ}
\end{equation} in order to estimate the true groundstate energy $E_0$. The preparation of the parametrized trial state $\psi(\bar{\theta})$ is done on the quantum computer, as well as the measurement of each Pauli string $\hat{P}_i$, while a classical computer is used to compose the complete value of the cost function $E(\bar{\theta})$ and optimize the set of parameters $\bar{\theta}$. The trial state is then updated with the newly calculated set of parameters, and used to generate the new circuit. In general, the wavefunction $\ket{\psi(\bar{\theta})}$ is obtained by applying a unitary transformation $U(\bar{\theta})$ to an initial reference state $\ket{\psi_{ref}}$. This reference can be either defined as the occupation of the spin-MOs according to the HF determinant, or generated by fixed circuit~\cite{tew2024}. The unitary transformation $U(\bar{\theta})$ is defined using a PQC.

\subsection{Quantum Mutual Information}
\label{ssec:qmi}
The Von-Neumann Quantum Mutual Information (QMI)~~\cite{vonneumann1996} is a property used to measure the correlation that can be found between two different components of a quantum system. 
Considering a system of $N$ elements for which exist a Hilbert space, $\mathcal{H}_u$, associated with each element $u \in \{1,\dots,N\}$, we can define the state of the system as the composition of these spaces. The resulting total space composition, denoted with $\ket{\Psi}$, belongs to the composed Hilbert space, $\mathcal{H}_{tot}$, obtained by the tensor product of each subsystem Hilbert space.

\begin{equation}
    \ket{\Psi} \in \mathcal{H}_{tot} = \bigotimes_{u=1}^{N}{\mathcal{H}_{u}}.
\end{equation}
To each quantum state $\ket{\Psi}$, we can define the associated \textit{density matrix} (or \textit{density operator)}
\begin{equation}
    \rho = \ketbra{\Psi}.
\end{equation}
Starting from a density matrix of a quantum state, we can retrieve information about only a subset of the elements, $k \subset B = \{1,\cdots,N\}$, defining the relative \textit{reduced density matrix} (RDM). The RMD of the subset $k$ is computed from the full density operator, $\rho$,  by tracing out the indices that are not the selected ones. We define $\bar{k}= B - k$ as the set of elements that have to be traced out. Thus, the RDM of the subset $k$, $\rho_{k}$, is formally defined as
\begin{equation}
    \rho_k = \text{tr}({\rho})_{\bar{k}}.
\end{equation}

\begin{figure}[!htb]
    \centering
    \includegraphics[scale=0.3]{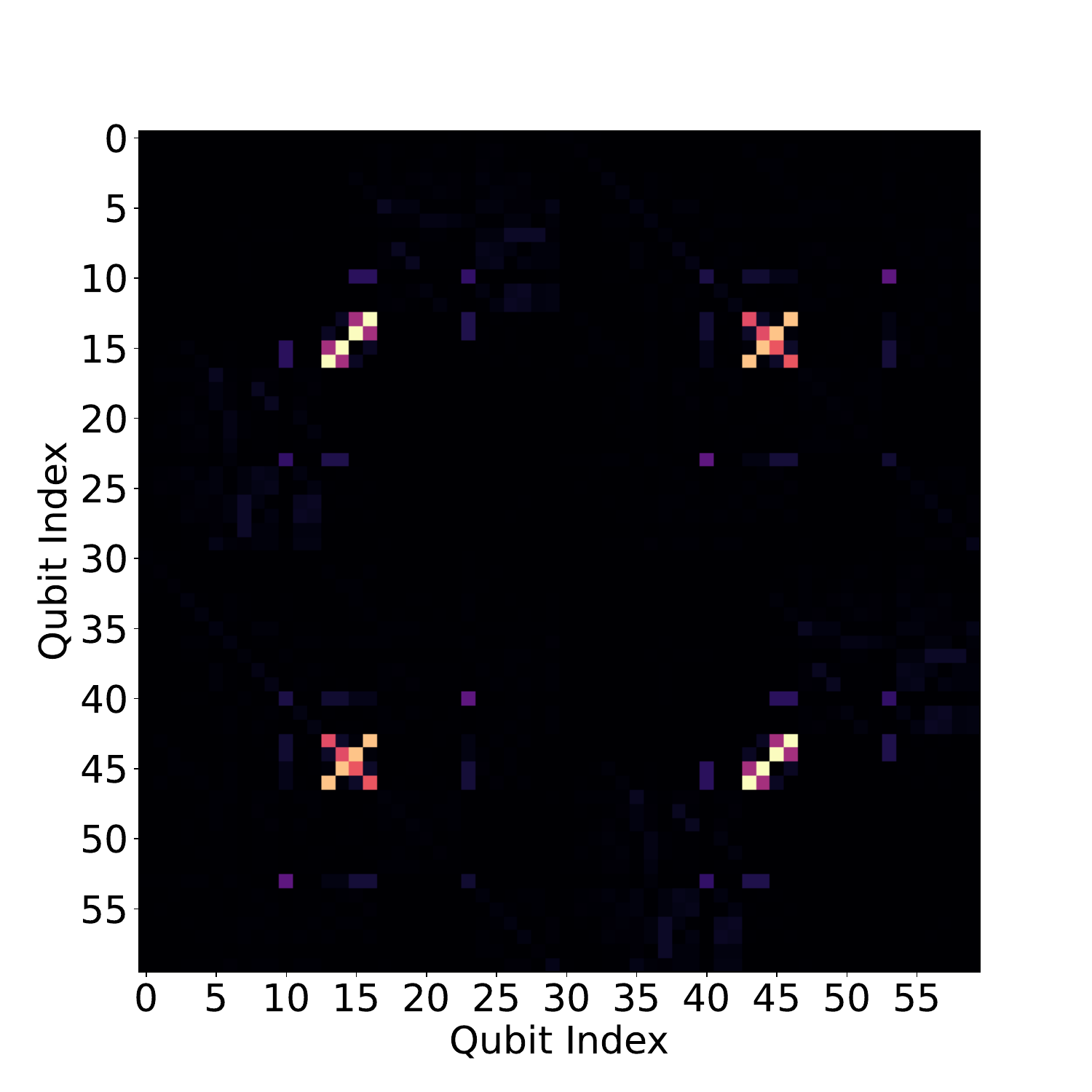}
    \caption{Example Quantum Mutual Information map of an $C_6H_6$ approximated wavefunction with CC-PVDZ basis and CAS(30,30). We used Jordan-Wigner mapping (spin-up orbitals from qubit indexes 0 to 29 and spin-down orbitals from qubit indexes 30 to 59).}
    \label{fig:exampleQMI}
\end{figure}

In order to compute the QMI matrix for the quantum system under study, for each component, $u$, and each pair of components, $(u,v)$, we need to compute the one and two element RDM respectively. 
Then, once all the reduced density matrices are obtained, we can compute the Von-Neumann entropy for each of them. The Von-Neumann entropy, $S$, is defined as 
\begin{equation}
\label{eqn:vnentopy}
    S(\rho) = -\Tr(\rho \log{\rho}).
\end{equation}
Following Equation (\ref{eqn:vnentopy}), we can define:
\begin{equation}
    S_{u} = S(\rho_u) = -\Tr(\rho_u\log\rho_u) 
\end{equation}
and
\begin{equation}
    S_{u,v} = S(\rho_{u,v}) = -\Tr(\rho_{u,v}\log\rho_{u,v}),
\end{equation}
where $\forall u,v \in \{1,\dots,N\}$.
Finally, from the above equations, we can define the \textit{Quantum Mutual Information Map} (or Matrix), $I$, which is formally defined as 
\begin{equation}
\label{eqn:QMI}
\begin{split}
    I_{u,v}&=(S_u + S_v - S_{u,v})(1 - \delta_{u,v})\\&=(S(\rho_u)+S(\rho_v)-S(\rho_{u,v}))(1-\delta_{u,v})
\end{split}
\end{equation}
where $u,v \in \{1,\dots,N\}$.
The QMI can be visualized with \textit{Quantum Mutual Information Maps} which are symmetric matrices with zero-valued entries on the diagonal by definition, shown in Figure \ref{fig:exampleQMI}. 
QMI found applications as a quantity to describe the electronic structure of a molecular system~\cite{Ding_2020}. This measure does not solely quantify the entanglement a system includes, but it takes in account both quantum and classical correlation. a value that describes only \textit{quantum correlation}.
We can consider the values of the $I$ matrix as a measure of the \textit{total} correlation between elements of the system, e.g. qubits, orbitals, or spin sites as used in this work. 

\subsection{Natural Orbitals}
\label{ssec:inos}
One of the possible way in which the complexity of the circuit is reduced is exploiting Natural Orbitals (NOs). In particular, we use them as one-electron basis set functions for both VQE simulations and QMI calculations.

Given a wavefunction $\Psi$, the associated NOs are defined as the set of molecular orbitals (MOs) for which the one-body reduced-density matrix (RDM) \begin{equation}
    \rho_{u,v} = \braketmatrix{\Psi}{a^{\dagger}_u a_{v}}{\Psi}
    \label{eqn:1b_rdm}
\end{equation}
is diagonal. From the diagonal terms of the RMD, suppose $\rho_{u,u}$, we can obtain information about the number of electrons in the $u$-th orbital and this value is denoted as Natural Orbital Occupation Number (NOON) of orbital $u$.

As claimed in~\cite{Lowdin1956}, in the NOs basis, the resulting CI expansion of the state under study is composed by the minimal number of Slater Determinants. The reduction of the population of SDs of a reference state is reflected in an increased sparsity of the Quantum Mutual Information map. The simplification of the variational problem by means of Natural orbitals in quantum computing has been studied in different works ~\cite{ratini2023, materia2023quantum, materia2024}.

For the \textit{recursive} nature of the basis i.e. the NOs depend on the wavefunction, which is itself defined by the NOs, constructing a CI wavefunction in this basis is tricky. To address this, an iterative procedure known as Iterative Natural Orbitals (INO)~\cite{Jafri1977} can be used, which aims to converge to MO where the wavefunction results in a diagonal one-body RDM. Achieving self-consistency with this iterative method is in principle costly, albeit the convergence rate is usually fast, so the process can be halted once a convergence criterion is achieved. 
\subsection{Retrieving QMI with SparQ}
\label{ssec:sparq}SparQ, or Sparse Quantum State Analysis, is an innovative tool designed to efficiently compute key quantum information theory metrics for wavefunctions that are sparse in their definition space~\cite{materia2024}. SparQ is particularly focused on wavefunctions derived from Post-Hartree-Fock methods and employs the Jordan-Wigner transformation to map fermionic wavefunctions into the qubit space. 
For excitation-based wavefunctions, this is done by applying excitations to the Hatree-Fock SD by the following equation:
\begin{equation}
    \label{eqn:exc-applied}
        \hat{a}_i^\dagger \hat{a}_j^\dagger \hat{a}_k \hat{a}_l |HF \rangle=\hat{a}_i^\dagger \hat{a}_j^\dagger \hat{a}_k \hat{a}_l \prod_{s \in HF} \hat{a}_s^\dagger |\varnothing \rangle,
\end{equation}
where the indexes $i,j \in Virtual$, $k,l\in Occupied$, and $s$ is the set of occupied orbitals that defines the Hartree-Fock SD. The fermionic to qubit mapping is then directly applied to the excitation operators.
This approach leverages the inherent sparsity of these wavefunctions to perform efficient quantum information analysis. This makes it possible to handle larger and more complex chemical systems than traditional methods such as the Density Matrix Renormalization Group (DMRG) even if sacrificing at times the quality of the wavefunction compared to the latter method. 

The sparsity of the wavefunction strictly depends on the Post-HF method in use, however, it is mostly related to these methods exploiting only a relatively low number of excitations in the overall Fock space.

Given a computational cost linear in the number of states, SparQ is able to handle any wavefunction given by the current methods, proving an invaluable tool for the aim of the present work.

\section{Multi-QIDA iterative method for shallow ansatz}
\label{sec:msmqida}
Multi-QIDA approach refines the initial QIDA~\cite{materia2023quantum} method by constructing ans\"atze through a multi-threshold procedure based on approximate QMI matrices, which enables more accurate and efficient variational quantum simulations. QIDA simplifies quantum circuit construction by using QMI to pre-determine where correlations between qubits are expected. This helps in creating circuits that are both efficient and effective for specific quantum problems, particularly in reducing the computational resources needed for VQE. Multi-QIDA applies the QIDA procedure multiple times to also incorporate mid-low correlations.
The application of the Multi-QIDA~\cite{tarocco2024} method demonstrates its utility in quantum simulations for strongly correlated lattice spin models like the Heisenberg model, as reported in a previous work~\cite {tarocco2024}. The procedure is composed of the following steps:
\begin{enumerate}
\begin{figure}[!htb]
    \centering
    \includegraphics[scale=0.45,trim={5cm 1cm 5cm 0},clip]{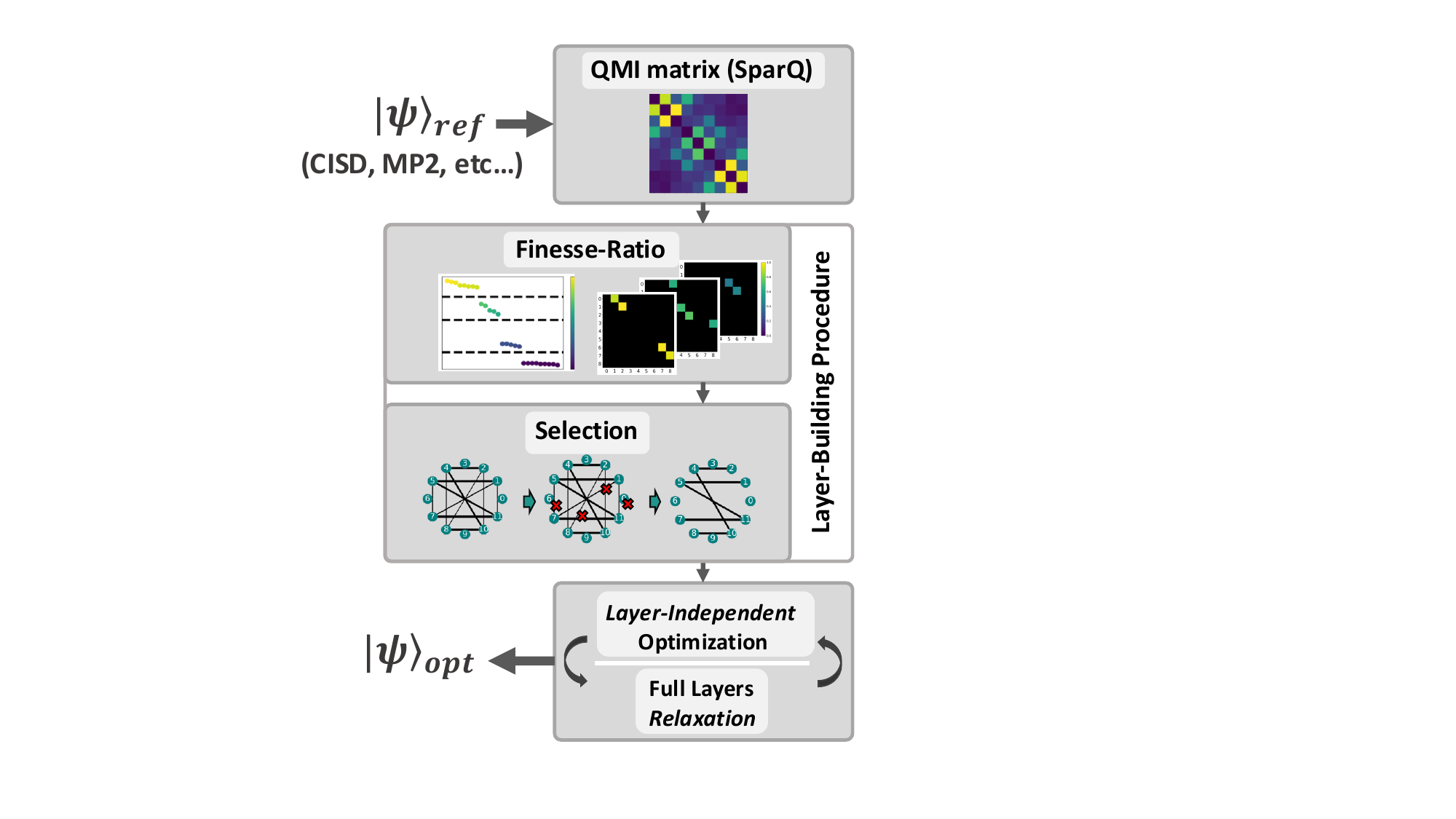}
    \caption{Multi-QIDA approach schematic workflow.}
    \label{fig:workflow}
\end{figure}
    \item \textit{QMI calculations}: The first step is to build the approximate QMI matrix exploiting the SparQ method, as defined in~\cite{materia2024}. (Section \ref{ssec:sparq})
    \item \textit{Layer-Building Procedure}: Without discerning between classical and quantum correlation, we spilt the qubit-pairs based on a selected set of QMI values, namely \textit{finesse-ratios}. For each range of QMI we obtain a QIDA-layer by performing a selection of only some relevant pairs. (Sections \ref{ssec:selection_crit} and \ref{ssec:lb})
    \item \textit{Layer-wise incremental VQE}: Each QIDA-layer is independently optimized, and then the full circuit goes through a \textit{relaxation} procedure. (Section \ref{ssec:iterVQE})
\end{enumerate}
The detailed workflow is shown in Figure \ref{fig:workflow}.

Key aspects of Multi-QIDA’s procedure are:
\begin{itemize}
    \item \textbf{Layered Ansatz Construction}: Multi-QIDA constructs variational layers step-by-step, where each layer is informed by the QMI matrix, selecting qubit-pairs based on their QMI values. This incremental addition of layers allows for the capture of crucial correlations that single-threshold approaches might miss.
    \item \textbf{Efficient Resource Management}: The algorithm is designed to reduce the computational overhead typically associated with ladder-style heuristic ansatzes, especially standard Hardware Efficient Ans\"atze. By selectively entangling qubits with strong correlations, Multi-QIDA reduced the number of required entangling gates, effectively constructing shallower circuits without sacrificing accuracy. This approach could be particularly valuable for current quantum hardware, where circuit depth and gate count directly impact performance due to noise.
    \item \textbf{Improved Convergence and Accuracy}: The iterative approach embedded in Multi-QIDA allows for faster convergence to the ground-state energy with fewer optimization runs. It consistently outperforms traditional ladder ansatz methods by maintaining high precision with reduced mean energy deviation, demonstrating its effectiveness in calculating ground-state energies accurately. Benchmarks showed improvement in both energetic terms and in accuracy compared to other ladder ansatz.
    
    \item \textbf{Mitigation of Barren Plateaus}: Following the idea that a multi-layer construction of the ansatz may create a funnel in the parameter space that can guide the minimization process, we may argue that Multi-QIDA’s iterative structure  may also increase the probability of avoiding barren plateau. Barren plateaus are a common issue in variational quantum algorithms where the optimization landscape becomes flat, complicating the parameter optimization process. By breaking down the variational landscape into manageable steps and refining parameters in stages, Multi-QIDA achieves better results compared to HEA ladder ansatz in which the full parameters space is defined since the beginning. 
\end{itemize}

\subsection{\textbf{SO}(4) as correlators}
\label{ssec:so4}
In the present work we focused only on the usage of more complex and expressible gates, instead of CNOTs, for the construction of the Multi-QIDA circuit.
The fully parametrized \textit{SO4} gate, from the \textbf{O}(4), group has been used in each QIDA-layer as an entangling gate. 
They offer a tunable correlation that can be chosen by the optimizer, as opposed to CNOTs-only-based ans\"atze in which the correlation is added similarly to a on/off switch.
Inside the possible transformation, a \textit{SO4} gate can also be parametrized to perform an identity, the \textit{fermionic-swap} gate, which allows to impose the fermionic anti-symmetrization to the ansatz, as well as Given rotations, which have been used in multiple Quantum Computing based Quantum chemistry application~\cite{Anselmetti_2021}. The identity plays an important role, since it can be the starting gate after the addition of a new layer. The choice of \textit{SO4} correlator is also justified as being the most general real-valued 4x4 matrices, performing only real transformation on the wavefunction, in line with the fact that the electronic Hamiltonian is composed only of real terms.

\begin{figure}[!htb]
    \centering
    \includegraphics[scale=0.3]{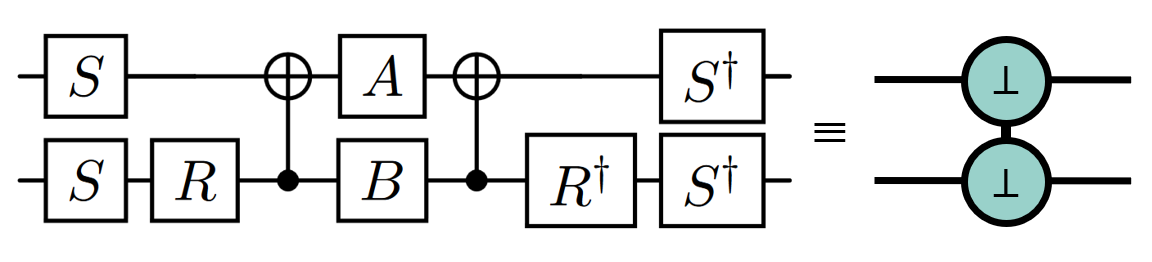}
    \caption{On the left, the circuit implementing a general \textit{SO4} gate. The gates A and B are general \textit{SU2} parametrized gates. On the right, the symbol we adopted in this work for such gate.}
    \label{fig:so4}
\end{figure}

A generic gate $U\in $\textbf{ SO}(4) is composed by two generic one-qubit rotations $A,B \in $ \textbf{SU}(2), four $S$ gates and two $R$ gates.
It is also known that every matrix $M \in$ \textbf{ SU}(2) can be written as a composition of $R_z(\alpha)R_y(\theta)R_z(\beta)$ for some $\alpha,\beta$ and $\theta$, while the $R$ gate is defined as $R_y(\pi/2)$ and the $S$ gate is obtained with $R_z(\pi/2)$. The $U$ gate is then parametrized using the two sets of three parameters of gates $A$ and $B$.

\subsection{Selection criteria}
\label{ssec:selection_crit}
As stated in~\cite{tarocco2024}, one of the main components of the construction of PQC following the Multi-QIDA method, is the request for a selection criterion for the reduction of the number of entangling pairs that fall into each layer. The objective of this procedure is to exploit cross-entanglement built among the Multi-QIDA layer in order to reduce the number of correlators.

In this work, we have exploited \textit{Minimum Spanning Trees }(mST) to select which correlators are going to be used in each Multi-QIDA layer. We start by defining a \textit{weighted graph}, $G=(V,E,w)$, where $V$ is the set of the vertices, $E \in V \times V$ is the set of the edges, and $w:E \rightarrow \mathbb{R}$ is a function that maps each edge $e_{i}\in E$ to a real valued weight $w_{e_{i}}$.
We can define a MST, $T\subseteq E$, which is a subset of the edges of the graph $G$ that satisfies the following properties:
\begin{itemize}
    \item \textit{Vertices Span}: The subset of the edges $T$ covers all the vertices $V$. Formally, for each pair of vertices $u,v\in V$, there exists a path in $T$ that allows to reach $v$ starting from $u$, and vice versa.
    \item \textit{Tree structure}: $T$ is an acyclic connected graph.
    \item \textit{Total Weight Minimizing}: Given the weight function $w$, the sum of the weight in the subset $T$ is minimized. Thus, if $T = \{e_1,e_2,\cdots,e_{|V| - 1}\}$, then the total weight $w(T) = \sum_{e\in T}w(e)$ is the lowest possible value among all the available spanning trees that can be built from the graph $G$.
\end{itemize}
In the same way, a \textit{Maximum Spanning Tree} (MST) can be defined following the previous properties by changing only the total weight values, which in this case, it has to be maximized. Thus, an MST is a subset $T\subseteq E$ that spans all the vertices $V$, it forms a tree, and the total weight $w(T)$ is maximized.

In this work, three different weight function has been used:

\textit{Maximum Correlation Spanning Tree (MCST)}: for each edge $e$ the associated weight $w_e$ is defined directly as the QMI value of the two vertices $v,u$ that they are connected by the edge $$w_e = w((u,v)) = I_{v,u}.$$ Thus, with this type of weight function, we are going to select the MST that collects the highest amount of correlation.
    \begin{center}\begin{figure}[!htb]
\centering
\begin{subfigure}[b]{.35\linewidth}
\centering
\includegraphics[width=\linewidth]{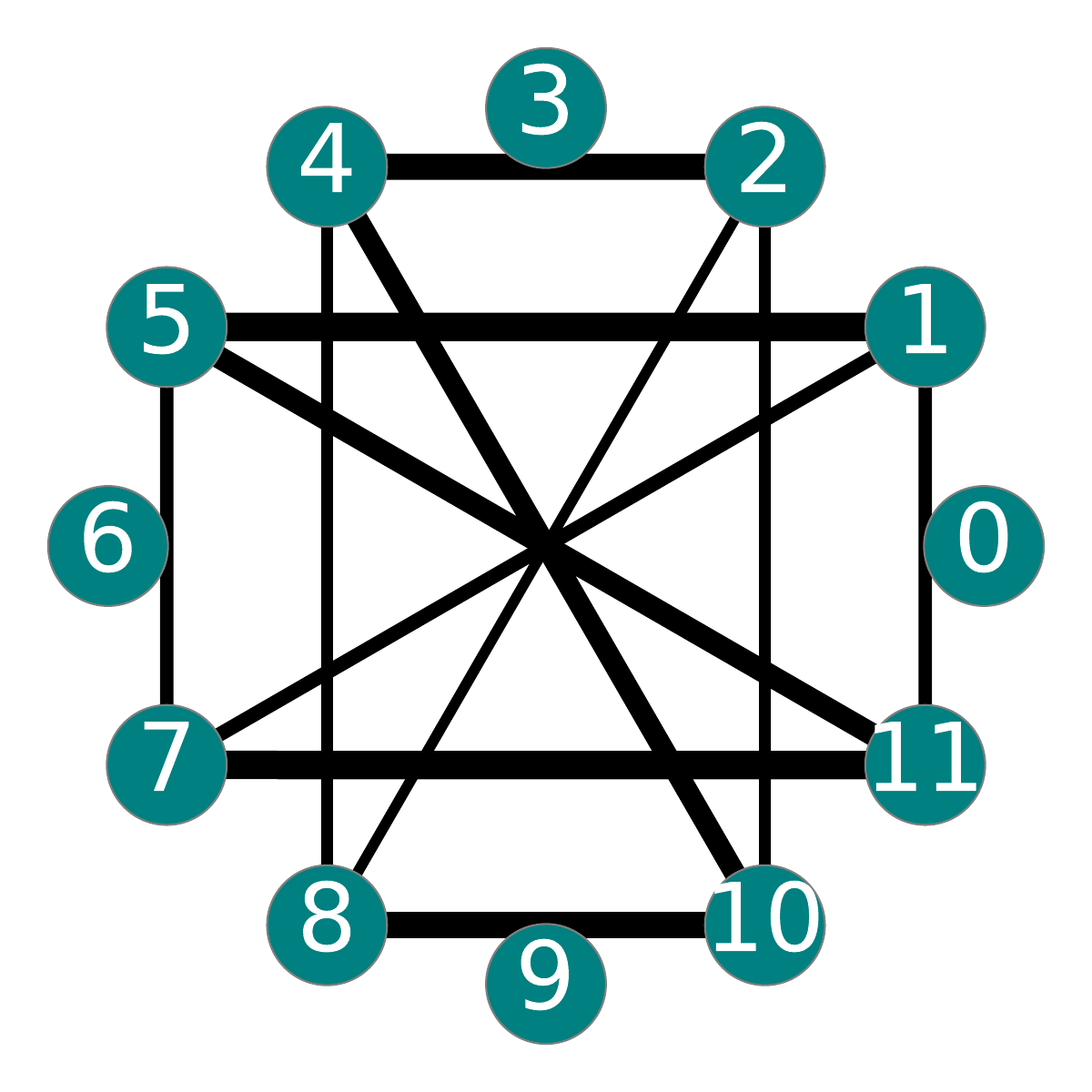}
\caption{}\label{sfig:max_no_red}
\end{subfigure}
\begin{subfigure}[b]{.35\linewidth}
\centering
\includegraphics[width=\linewidth]{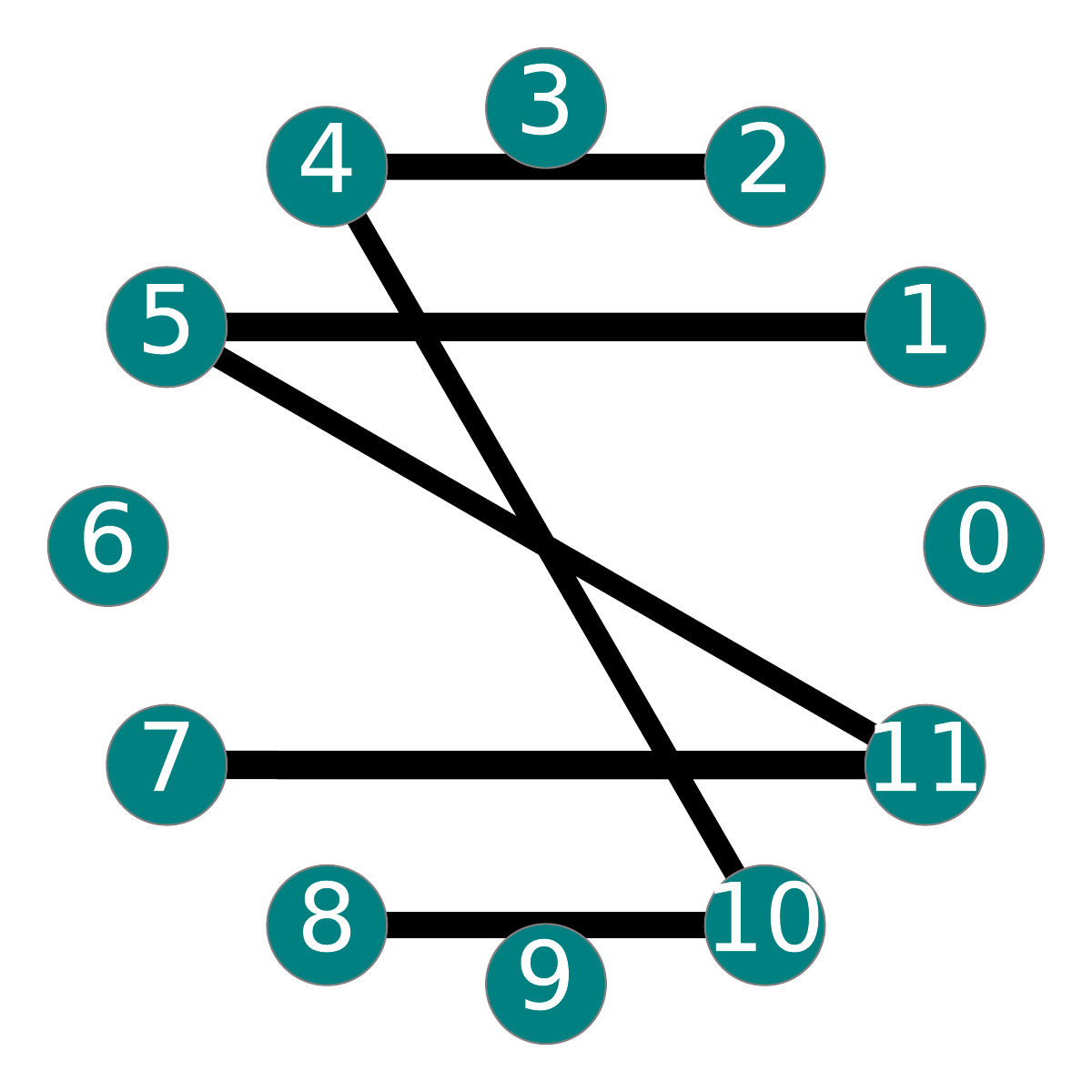}
\caption{}\label{sfig:max_red}
\end{subfigure}
\caption{Pictorial representation of the action of the correlators reduction. The width of each edge corresponds to the value of QMI between the two qubits, $I_{uv}$. In the two figures: (a) Original layer without any reduction applied. (b) Maximum correlation reduction applied leading to a QIDA-layer composed of the remaining edges.}
\label{fig:full_to_max}
\end{figure}
\end{center}

\textit{Distance Reduction Spanning Tree} (DRST): For each edge $e$, the relative weight is defined as $$w_e = w((u,v)) = d(u,v),$$ where $d(u,v)$ is a \textit{topology-based distance} function. The distance term is defined as the topological distance between qubits $u$ and $v$,  which can be seen as the number of edges that are included in the shortest path to connect these two qubits. From another point of view, this distance $d(u,v)$ can be defined as the minimum number of 2-qubits SWAP gates needed to make qubits $u$ and $j$ next-neighbour. For this specific case, the topology is linear, thus the distance function is defined as $|u - v|$. This weight function implemented on linear topology corresponds to the \textit{empirical reduction} used in~\cite{materia2023quantum}, i.e. the objective is to consider qubit-pairs as close as possible to the diagonal. Thus, it is required to build a Minimum Spanning Tree.
    \begin{center}
        
    \begin{figure}[!htb]
\centering
\begin{subfigure}[b]{.35\linewidth}
\centering
\includegraphics[width=\linewidth]{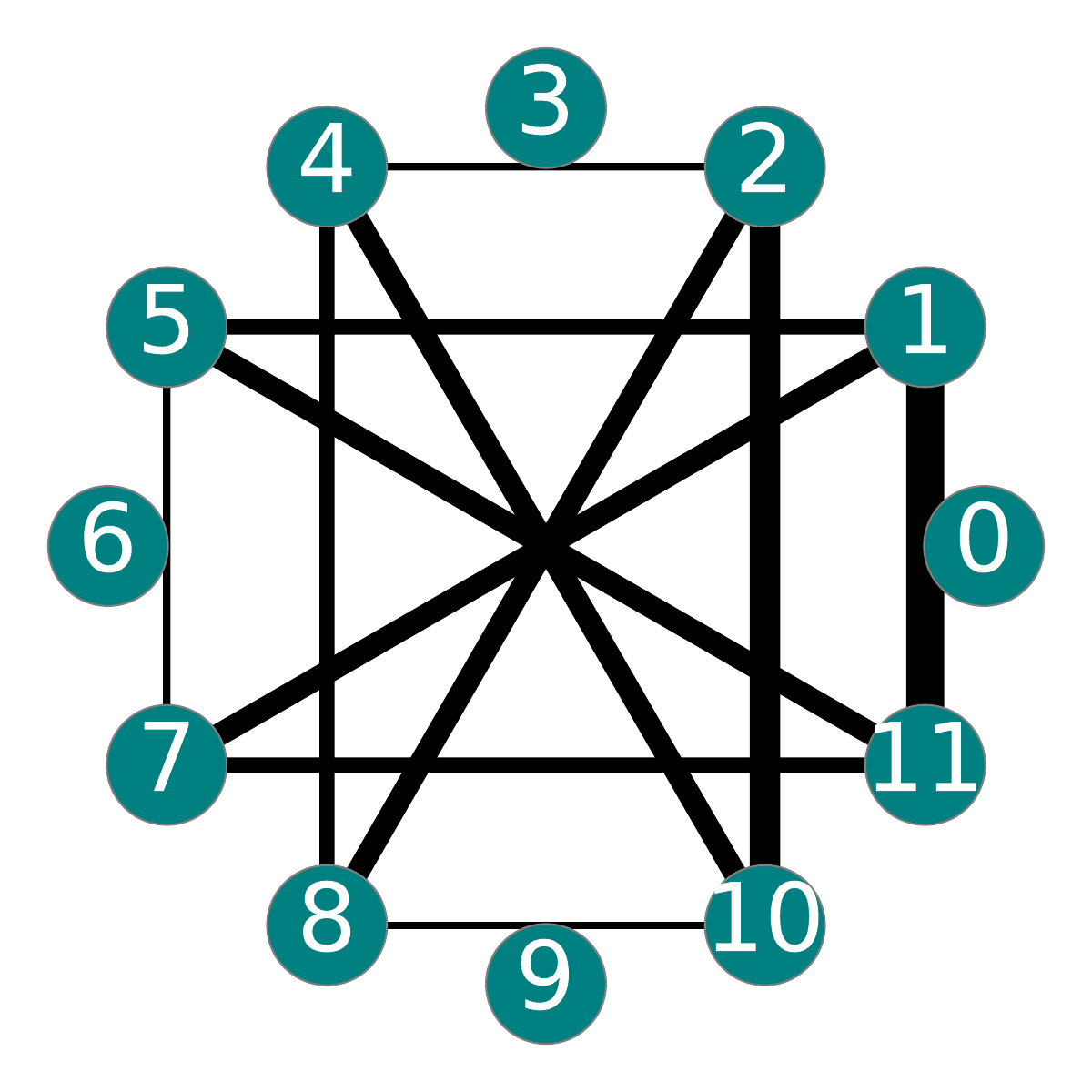}
\caption{}\label{sfig:emp_no_red}
\end{subfigure}
\begin{subfigure}[b]{.35\linewidth}
\includegraphics[width=\linewidth]{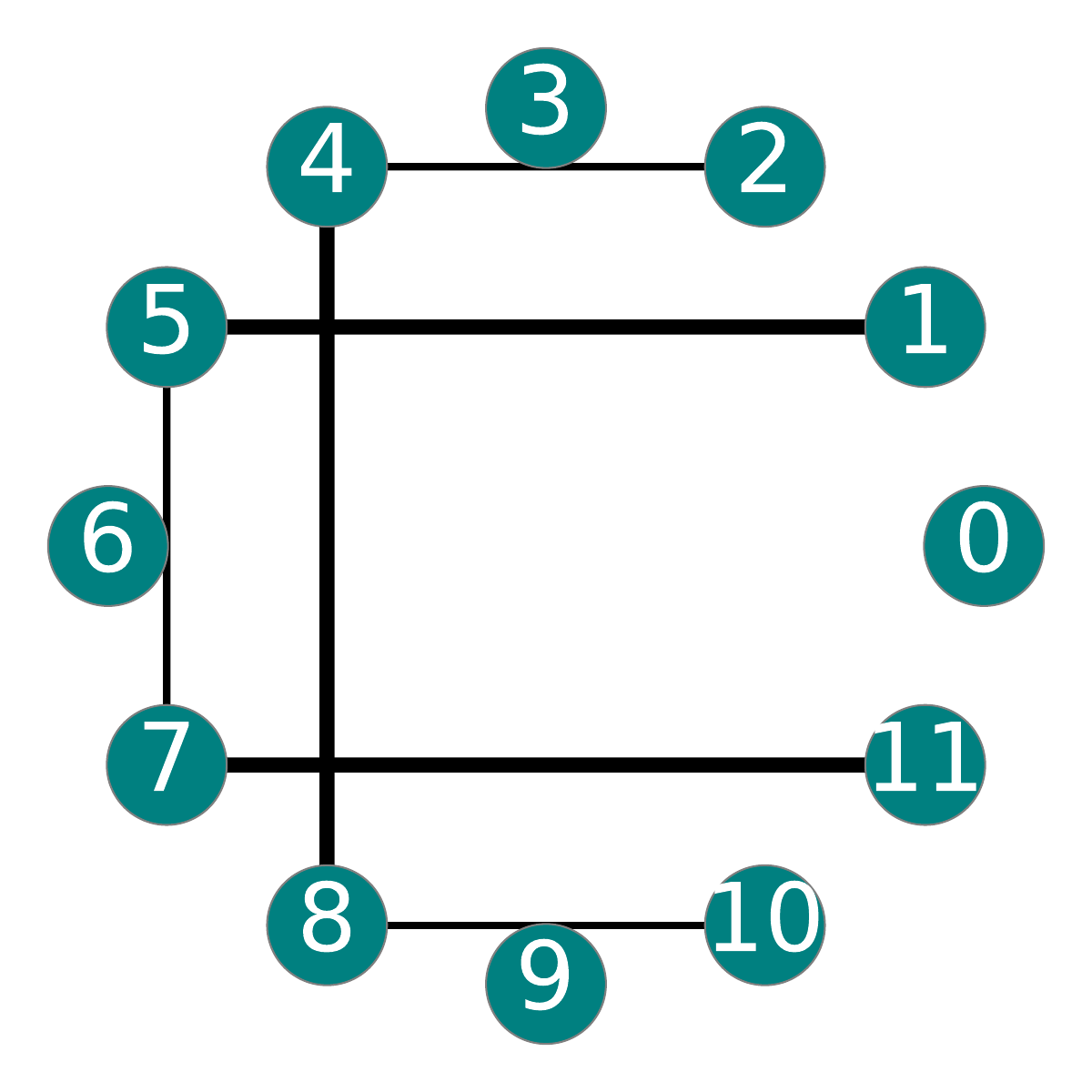}
\caption{}\label{sfig:emp_red}
\end{subfigure}\caption{Pictorial representation of the action of the correlators reduction. The width of each edge corresponds to the topological distance between the two qubits, $d(u,v)$. In the two figures: (a) Original layer without any reduction applied. (b) Empirical reduction leads to a QIDA-layer composed only of the identified edges.}
\label{fig:full_to_empirical}
\end{figure}
\end{center}

In Figures \ref{fig:full_to_max} and \ref{fig:full_to_empirical}, two examples of correlation reduction are shown. In particular, they represent how a candidate set of entangling gates, image (a) in both figures, gets reduced by the application of the selection criteria. The weights associated with each edge are defined randomly to be general.
Exploiting these two selection criteria, in the next section, we illustrate in pseudo-code the algorithm used to construct each Multi-QIDA layer.

\subsection{Layer-building construction}
\label{ssec:lb}
As explained in the previous section, our goal is to create a shallow-depth circuit for state preparation, improving the results obtained by the original QIDA method, in order to include a wider spectrum of different qubit-pair correlation.
In this section, we will briefly define the pseudo-code to define each layer, which is a slight modification of the one presented in~\cite{tarocco2024}.

\begin{figure}[!htb]
    \centering
    \includegraphics[scale=0.3]{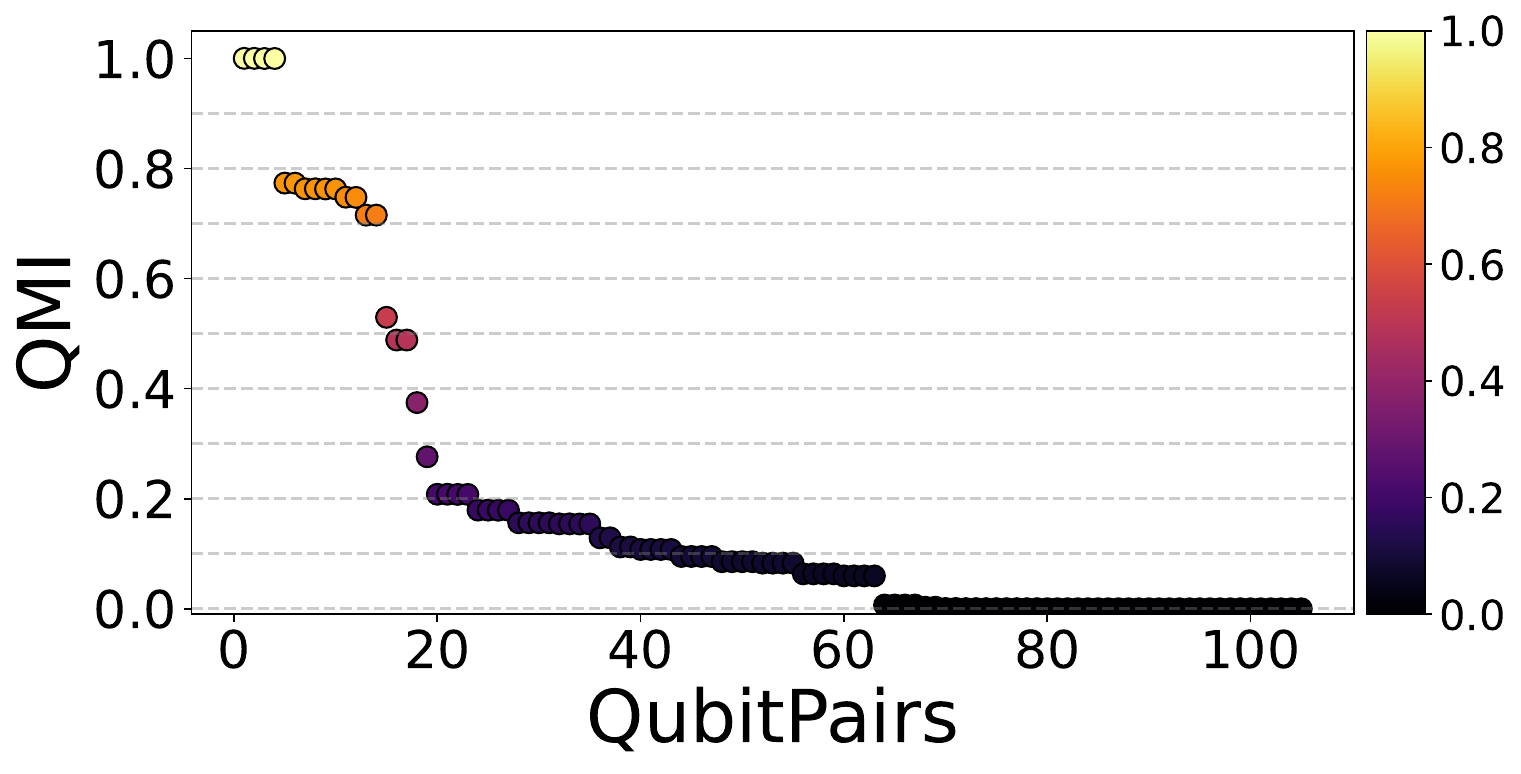}
    \caption{Example of the decreasing values of QMI for NH$_3$ INOs molecular system.}
    \label{fig:qubit_pairExample}
\end{figure}

The selection is carried out on the list of descending QMI-value order qubit-pairs obtained from the QMI matrix $I_{u,v}$. Using a threshold $\mu$, called \textit{finesse-ratio}, we can decide the size of the chunks of qubit-pairs that a Multi-QIDA layer has to contain.
The finesse-ratio is not computed automatically with a fixed step, it is instead empirically determined based on the distribution of the qubit-pairings. By observing the distribution of the QMI spots, it can be noticed that it decays rapidly, concentrating on a minor group of highly/mid correlated spots (as Shown in Figure \ref{fig:qubit_pairExample}), and thus, as the lower the interval of correlation is chosen, the higher the number of pairings included in the chunks.

In section \ref{ssec:qmi_matrix}, the finesse-ratios used for each system are shown, for explanation purposes only, we are going to call $\bar{\mu}$ the list of finesse-ratios used to create a Multi-QIDA ansatz.

\begin{algorithm}[!htb]
\caption{Schematic outline of the Multi-QIDA Layers-builder}
\begin{algorithmic}[1]
\Require $I_{ij}$, $\bar\mu$, $N_{qubits}$, $w(\cdot, \cdot)$
\Ensure List of entangling map $L$
\State $L \gets$ empty list
\State $m\gets 0$
\For{$m \in [0,\cdots,len(\bar\mu)-1]$}
\State $G \gets  (V=\{1,\cdots,N\}, E=\emptyset)$
\For{$q_u,q_v \in \{\forall I_{u,v}: \bar\mu[m]>I_{u,v}\geq \bar\mu[m+1]\}$}
\State $G.add\_edge(q_u,q_v, w(u,v))$
\EndFor
\State $T \gets ComputeMST(G)$
\State $L.append(T.get\_edges())$
\EndFor
\State \textbf{return} $L$

\end{algorithmic}
\label{algo:sel_pseudocode}
\end{algorithm}
For each pair of consecutive finesse ratios, we define the range, $l$, in which the qubit-pairs are used to impose an edge on a graph $G_l$. This range corresponds to a Multi-QIDA layer. Selected all the pairings that fall in the selected QMI value range, for each $u,v$ in the chunk, we define a weighted edge that connects vertex $u$ to vertex $v$, with weight $w(u,v)$. The weight function $w(\cdot,\cdot)$ depends on the selection criteria according to the one defined in the previous section (Section \ref{ssec:selection_crit}). Once all the edges are inserted in the graph, we create the MST (or mST), $T_l$, concerning the cost function used to reduce the number of correlators. From $T_l$, we can now retrieve the collection of the edges $E(T_l) \subseteq E(G_l)$ that will compose the entangler map for the $l$-th layer.

At each step, the graph  $G_l$ is reset, ensuring that only the correlators contained within a given chunk are considered. This guarantees that the minimum spanning tree (MST) is constructed solely from the correlators relevant to that chunk, thereby preventing any edges from previous layers from being reused or exploited.

For each tree  $T_l$, the list of edges forming the MST (denoted as mST) is translated into a QIDA-layer. Once all such layers are generated, a final "ladder layer" is appended. This layer arranges the remaining correlators in a top-down topology, allowing separate correlation groups to be interconnected. Functionally, the ladder layer acts as a final selection stage for qubit pairs that fall outside the specified finesse-ratio range. By applying MST-based selection to these leftover pairs, the resulting topology of this layer approximates that of a ladder both in term of structure and in CNOT cost.

All QIDA-layers and the final ladder layer are then assembled into a complete set of layers, denoted as $L$ . This collection $L$  is subsequently passed to the layer-wise iterative VQE algorithm, which uses it as a blueprint to construct the adaptive quantum circuit.

\subsection{Incremental VQE optimization}
\label{ssec:iterVQE}
%\red{\sout{As performed in the previous work, we defined the optimization procedure }}
To optimize the full Multi-QIDA circuit, the optimization is defined as an incremental routine with the aim to not starting with a complete and complex circuit, but instead adding and optimizing only one QIDA-layer at a time.

The selection of a suitable optimization routine is necessary when dealing with different circuit layouts in which the composition of the ansatz may vary from layer to layer. As the name of this section suggests, this will be done iteratively along many steps, each of which will include two main phases, an optimization of the single layer $L_l$ at the $l$-th step, and a global optimization of all the previous layers $\{ 0 \dots, l-1 \}$. We remind the reader that each layer $L_l$ is composed of \textit{SO4} gates as correlators; however, the procedure explained here is independent of this variable. 

Different from our previous work ~\cite{tarocco2024}, for these molecular systems, the initialization to the identity of the additional layers turned out not being the right choice, due to the presence of local minima, preventing the optimization from proceeding. 
To overcome these limitations, we decided to initialize the new layer using an operator that is close to identity but not identity. This goal is achieved by adding additional layers which has a random offset from the identity. The set of initial parameters for the additional layer, $\bar\theta_{l}^i$, is created by randomly sampling values from a uniform distribution with mean $0$ and standard deviation $0.1$, i.e. $\bar\theta_{l}^i \in_R \mathcal{U}(0,0.1)$. The value 0.1 has been empirically estimated by selecting the lowest value that allowed escaping the local minima of the previous QIDA-layer, while recovering in a few VQE iterations the previous energy after the addition of the successive layer. 
Now, we can briefly define the two steps required to optimize the circuit after the addition of the $l$-th QIDA-layer: We first perform an independent optimization of the unitary transformation $U_l(\bar{\theta}_l^i)$, that add the new layer to the previous solution. The action of the unitary is defined as $\ket{\Psi_{l}}= U_l(\bar{\theta}_l^i)\ket{\Psi_{l-1}}$, where $\ket{\Psi_{l-1}}$ is the previous solution state. The initial parameters are sample as previously indicated, and after the optimization, we obtain the set of optimal parameters for the $l$-th QIDA-layer, denoted with $\bar{\theta}^{i*}_l$. 
We then perform a relaxation of the full circuit. The relaxation is defined by \begin{equation}
    \langle\Psi_0|[\prod_{j=l}^{j=0}U_j^\dagger](\bar{\theta}_l^r) | \mathbf{H} |[\prod_{j=0}^{j=l}U_j](\bar{\theta}_l^r)|\Psi_0\rangle,
\end{equation}
where $\ket{\Psi_0}$ is the initial or reference state, $\mathbf{H}$ is the Hamiltonian in the qubit space, the product of unitaries $[\prod_{j=0}^{j=l}U_i]$ is the empty circuit up to the $l$-th layer, and $\bar{\theta}_l^r=\bar{\theta}_{l-1}^{r*} + \bar{\theta}_l^{i*}$ is the concatenation of optimal parameters of the relaxation procedure from the previous layer, $l$-1, and the optimal parameters of the independent optimization of layer $l$. After the relaxation procedure, the set of optimal parameters up to the layer $l$ is defined and denoted with $\bar{\theta}_l^{r*}$.

To briefly resume a general procedure that is executed at each step $l$, exploiting non-fixed parameters quantum circuit, i.e. $QC_{empty}$:
\begin{enumerate}
    \item Append the $l$-th layer to the $QC_{empty}$;
    \item Assign the optimal parameters up to the previous iteration $\bar{\theta}_{0,1,\dots,l-1} = \bar{\theta}^*_{prev}$;
    \item Initialize the parameters of the $l$-th layer to an offset of the identity;
    \item Find the optimal parameters of the $l$-th layer alone, $\bar{\theta}^*_{l}$;
    \item Compose the total set of parameters up the the $l$-th layer as $\bar{\theta}_{tot} = \bar{\theta}_{prev}^*+\bar{\theta}_{l}^*$;
    \item Use $\bar{\theta}_{tot}$ as starting parameters for a final VQE in which the variational wavefunction is defined by $QC_{empty}$;
    \item Once converged, the optimized set of the combined circuit, i.e. $\bar{\theta}_{tot}^*$ is obtained.
\end{enumerate}

\begin{algorithm}[!htb]
\caption{Iterative (Re)-Optimization routine}
\begin{algorithmic}[1]
\Require $N_{qubits} > 0$, List of entangling map $L$.
\Ensure Optimal parameters $\bar{\theta}^*_{tot}$, Converged energy $E_{tot}$
\State $QC_{empty} \gets $QuantumCircuit($N_{qubits}$)
\For{$l \in L$}
\State $add\_l \gets$True
\While{$add\_l$}
\State append($QC_{empty},l)$
\If{$l=0$ }
    \State $\bar\theta_{0}\in_R[0,2\pi)$ 
    \State $E_0,\bar{\theta}_{tot}^* \gets$VQE$(QC_{empty},\bar{\theta}_0)$
\Else
\State $\bar{\theta}_{l} \in_R \mathcal{U}(0,0.1)$
    \State $E_l,\bar{\theta}_{l}^{*} \gets $VQE$(QC_{prev}, \bar{\theta}_{l})$
\State $\bar{\theta}_{tot} \gets \bar{\theta}_{prev}^* + \bar{\theta}_{l}^{*}$
\State $E_{tot},\bar{\theta}_{tot}^* \gets $VQE$(QC_{empty}, \bar{\theta}_{tot})$

\EndIf
\State $\bar{\theta}_{prev}^* \gets \bar{\theta}_{tot}^*$
\State $QC_{prev} \gets $assign$(QC_{empty},\bar{\theta}_{prev}^*)$
\EndWhile
\EndFor
\Return $E_{tot}, \bar{\theta}_{tot}^*$
\end{algorithmic}
\label{algo:opt_layer}
\end{algorithm}
We notice that the optimization procedure shown in Algorithm \ref{algo:opt_layer} is not limited to the use in combination with Multi-QIDA, but could also be used as an optimization method for other iterative ans\"atze, as it is a similar procedure to the one used by the Adapt-VQE algorithm~\cite{Grimsley_2019}, as well as in Layer-VQE~\cite{Liu_2022}.

\section{Computational details}
\subsection{Molecular Systems}
\label{ssec:systems}
To test the Multi-QIDA approach on molecular systems, we considered five different molecules that, once codified on the quantum computer, span the range between 8 to 14 qubits.
The systems chosen are H$_2$O, NH$_3$, and BeH$_2$ in terms of full-size system, thus Full-CI level,  while H$_2$O and N$_2$ with bigger basis have been studied at CASCI level.
In Table \ref{tab:mol_system}, we have summarized all the information related to the system under study.
\begin{table}[!htb]
    \centering
    \fontsize{8pt}{10pt}\selectfont
    \setlength\tabcolsep{4pt}
    \begin{tabular}{|c||c|c|c|c|}
    \hline
        Mol.& Coordinates(\AA) &Basis&Qubits\\
        \hline\hline
         H$_2$O& \makecell{$H$ $0.757$ $0.586$ $0.0$ \\$H$ -$0.757$ $0.586$ $0.0$\\$O$ $0.0$ $0.0$ $1.595$}&\makecell{INOs\\RCISD\\(STO-3G)}&  12\\\hline
         BeH$_2$ &\makecell{$Be$ $0.0$ $0.0$ $1.334$\\$H$ $0.0$ $0.0$ $0.0$\\$H$ $0.0$ $0.0$ $2.668$} &\makecell{INOs\\RCISD\\(STO-3G)}& 12 \\\hline
         NH$_3$& \makecell{$N$ $0.0$ $0.0$ $0.1211$\\ $H$ $0.0$ $0.9306$ -$0.2826$\\ $H$ $0.8059$ -$0.4653$ -$0.2826$\\ $H$ -$0.8059$ -$0.4653$ -$0.2826$}&\makecell{INOs\\RCISD\\(STO-3G)}& 14\\\hline\hline
         \makecell{H$_2$O\\ CAS(4,4)} & \makecell{$H$ $0.847$ $0.0$ $0.0$ \\$H$ -$0.298$ $0.0$ $0.793$ \\$O$ $0.0$ $0.0$ $0.0$}&6-31G&  8\\ 
         \hline
         \makecell{N$_2$\\ CAS(6,6)} & \makecell{$N$ $0.0$ $0.0$ -$0.5488$\\$N$ $0.0$ $0.0$ $0.5488$}&cc-pVTZ& 12\\\hline

    \end{tabular}
    \caption{Molecular systems under analysis. The number of qubits is computed as 2*$M_{act}$, where $M_{act}$ are the active orbitals. For the first three system, the basis-set in brackets is the initial one on which INOs are constructed.}
    \label{tab:mol_system}
\end{table}
Only the frozen core approximation at the Hartree-Fock level has been used for the first group of three molecules, by freezing the first core orbital for each system.
All the systems in the first group have been analyzed with STO-3G basis set. One HF is computed, we have applied the procedure to obtain INOs for each system, starting from RCISD calculations. The INOs obtained are used as set of MOs that will define the true active space of the system, i.e. $M_{act}$.

For the second group of molecules instead, a more fine selection is performed. For H$_2$O, the CAS(4,4) active space includes 4 electrons in 4 orbitals: two $\sigma$ bonding (O-H) and two $\sigma^*$ antibonding orbitals. Using the 6-31G basis set provides a moderate description of the molecular orbitals, accurately representing the bonding (HOMO) and antibonding (LUMO) levels. For this system, the selected orbitals will compose the set $M_{act}$ with size 4. For N$_2$, the CAS(6,6) active space involves 6 electrons distributed across 6 orbitals: bonding $\sigma_g(2p_z)$, $\pi_u(2p_x)$, $\pi_u(2p_y)$, and antibonding $\sigma_u^*(2p_z)$, $\pi_g^*(2p_x)$, $\pi_g^*(2p_y)$. For the $N_2$, the set of $M_{act}$ will be composed of 6 elements. Using the cc-pVTZ basis set, the orbital energies are more accurate due to better flexibility and polarization functions, resulting in a realistic HOMO-LUMO gap. In this case, the HOMO is degenerate, including both the $\pi_u(2p_x)$ and $\pi_u(2p_y)$  orbitals, as well as the LUMO, composed by $\pi^*_u(2p_x)$ and $\pi^*_u(2p_y)$. In contrast, the STO-3G basis set often predicts a different ordering, where the $\sigma_u^*(2p_z)$ orbital can incorrectly appear lower in energy than the $\pi_g^*$, leading to an underestimated HOMO-LUMO gap. 
With cc-pVTZ, the active space selection better captures the electron correlation essential for describing the triple bond in N$_2$. In minimal sets like STO-3G, improper orbital energy ordering may lead to inaccuracies in multiconfigurational calculations~\cite{casscf_n2}.

Starting from the set of $M_{act}$, we can define the second quantization Hamiltonian $\hat{H}$ (Equation \ref{eq:H_elec}), applying Jordan-Wigner mapping to obtain the electronic molecular Hamiltonian in the qubit space, $\hat{\mathbf{H}}, $ as defined in Equation \ref{eqn:H_elec_qubit}. The number of qubit related for each system will be $2\times M_{act}$, split between $|M_{act}|$ spin-$\alpha$ qubits and $|M_{act}|$ spin-$\beta$ qubits, with ordering $\ket{\alpha\dots\alpha\beta\dots\beta}$.
\subsection{Heuristic Ans\"atze Comparison}
\label{ssec:hea}
We decided to compare our approach with the most general variational wavefunction as the Hardware-Efficient Heuristic ansatz. 
In particular, the way in which the correlators are placed for this type of ans\"atze is in ladder fashion: a sequence of rotation gates is followed by a series of CNOTs placed in a top-ordering connecting adjacent qubits, this configuration repeated $d$ times, and completed by a final series of parametrized rotation gates. The depth of the ladder has been chosen to be of the order of magnitude of Multi-QIDA CNOT counts. The number of CNOTs for ladder fashion circuits is defined as $(N$-$1)*d$, where $N$ is the number of spin-orbitals, and $d$ is the depth. HEA are denoted with $(L)^{CX}_{5}$, while Multi-QIDA using the label $QIDA_{sel}$, where $sel$ can be one of the two selection criteria defined in Section \ref{ssec:selection_crit}. In particular, the complete CNOTs count is shown in Table \ref{tab:cnots}.
\begin{table}[!htb]
\centering
    \fontsize{7pt}{10pt}\selectfont
    \setlength\tabcolsep{4pt}
    \begin{tabular}{|c||ccc|cc|}\hline
         \#CNOTs& BeH$_2$ &H$_2$O&NH$_3$&\makecell{H$_2$O\\CAS(4,4)}&\makecell{N$_2$\\CAS(6,6)}\\\hline\hline
         $(L)^{CX}_{d}$& 66 & 55 & 65 & 35 & 66 \\\hline
         $QIDA_{max}$&70&58&66&36&68\\\hline
         $QIDA_{emp}$&70&58&66&36&68\\\hline
    \end{tabular}
    \caption{Number of CNOTs used by each ansatz configuration. $(L)_d^{CX}$ denotes HEA, while $QIDA_{sel}$ to different Multi-QIDA ans\"atze. The number of CNOTs in Multi-QIDA configuration is equal due to the selection based on spanning trees.}
    \label{tab:cnots}
\end{table}

\subsection{Metrics and Measures}
\label{ssec:mm}
The main metric used to compare different ansatz configurations is the number of CNOTS, \#CNOT. We decided to not employ the measurement of the depth of the circuit, due the fact that for the Multi-QIDA method, the full circuit is composed by entangler maps that differs from layer to layer. Thus, a different number and disposition of the correlators,  leads to an inhomogeneous metric. 
To measure the performance of the variational calculation, we used used \textit{percentage correlation energy}, $\epsilon$, defined as
\begin{equation}
    \begin{split}
    \epsilon_i &= 100 \cdot \frac{E_{{VQE}_i} - E_{HF}}{E_{FCI} - E_{HF}}\\
    &=100 \cdot \frac{\langle \psi(\bar{\theta})_i|\mathbf{\hat{H}}| \psi(\bar{\theta})_i\rangle - E_{HF}}{E_{FCI} - E_{HF}}
    \end{split}
    \label{eqn:ecorr}
\end{equation}
where $E_{VQE_{i}}$ is the converged energy of the $i$-th simulation, $E_{HF}$ is the Hartree-Fock SCF energy, while $E_{FCI}$ is the exact solution, obtained by performing a diagonalization on the qubit Hamiltonian defined on the $M_{act}$ orbitals, and selecting the lowest eigenvalue. The $E_{HF}$ is instead directly obtained by the RHF solver of PySCF~\cite{Sun2018, Sun2020, pyscf} package.

For the system in which a specific active space is selected i.e. H$_2$O-CAS(4,4) and N$_2$-CAS(6,6), the reference exact energy correspond to the CASCI energy, $E_{CASCI} =E_{core} + E^{diag}_{act}$, where $E_{core}$ is the energy contribution of inactive occupied orbitals, and $E^{diag}_{act}$ is the lowest eigenvalue of the active Hamiltonian. Equation \ref{eqn:ecorr} can be redefined considering an active Hamiltonian as
\begin{equation}
\epsilon_{act} = 100\cdot\frac{E_{VQE_i} - E_{HF}}{E_{CASCI} - E_{HF}}.
    \label{eqn:ecorr_act}
\end{equation}

We have then computed the \textit{Mean Correlation Energy Deviation} (MCED) which quantifies the average deviation of the correlation energy from the best performing simulation $\epsilon_{best}$. It is obtained by summing up the difference between the correlation energies of each VQE simulation and the correlation energy of the best performing simulation, then, the sum is normalized by the total number of simulations. The MCED is formally defined as
\begin{equation}
MCED  = \frac{\sum^{\#VQEs}_{i}{|\epsilon_i - \epsilon_{best}|}}{\#VQEs},
    \label{mced}
\end{equation}
where $\epsilon_i$ is the correlation energy for a specific VQE run,  $\epsilon_{best}$ is the correlation energy of the best performing simulation, and $\#VQEs$ is the total number of simulations for a given ansatz configuration. All the results related to average and best-performing simulations for each ansatz configuration are collected in Table \ref{tab:ino_systems_energies_main} for INOs systems, and in Table \ref{tab:cas_systems_energies_main} for Active-Space systems.

We also computed four more quantities for each wavefunction computed by both Multi-QIDA and HEA circuits. We are interested in :\begin{itemize}\item \textbf{Fidelity with the true ground-state}: \begin{equation}
\mathcal{F} = \langle\Psi_{GS} | \psi({\bar{\theta}})_i\rangle,
\end{equation} where $i$ is the $i$-th VQE simulation.\item\textbf{Projection along z-axis of the spin}:
\begin{equation}
\begin{split}
    \hat{S}_z &= \frac{1}{2}(\hat{N}_{\alpha} - \hat{N}_{\beta})\\&= \frac{1}{2}(\sum_i^{M_{act}}a_{i,\alpha}^\dagger a_{i,\alpha}^{} - \sum_i^{M_{act}}a_{i,\beta}^\dagger a_{i,\beta}^{}).
\end{split}
\end{equation}\item\textbf{Spin squared} which can be defined as
\begin{equation}
    \hat{S}^2 = \hat{S}_{-}\hat{S}_{+} + \hat{S}_z(\hat{S}_z + 1),
\end{equation} where $$\hat{S}_{-} \text{=} \sum_{i}^{M_{act}}a_{i,\beta}^{\dagger}a_{i,\alpha}^{} \text{ and } \hat{S}_{+}\text{=}\sum_{i}^{M_{act}}a_{i,\alpha}^{\dagger}a_{i,\beta}^{},$$
\item\textbf{Number of particles} in this case, electrons, obtained from
\begin{equation}
    \hat{N}_e = \sum_{\sigma\in\{\alpha,\beta\}}\sum_{i}^{M_{act}}a_{i,\sigma}^{\dagger}a_{i,\sigma}^{}.
\end{equation}
\end{itemize}
All the four additional properties measured, the values of best-performing results, are collected in Table \ref{tab:ino_systems_props_main} for INOs systems, and in Table \ref{tab:cas_systems_props_main}  for Active-Space systems.

\subsection{Simulation details}
\label{ssec:sim_det}
The part of classical Quantum Chemistry included in this work is performed with the PySCF Python package. The computation of the INO is made using Restricted-CISD (RCISD).
We used the QuAQ (Quantum@L'Aquila)~\cite{QuAQ2025} code base for most of the preprocessing part, as well as for the definition of the Iterative VQE procedure and the Multi-QIDA layer builder. For the computation of the QMI map from the reference wavefunction, we used SparQ ~\cite{materia2024} algorithm, which is actually contained in QuAQ codebase.
For the creation of quantum circuit and anything directly related with them, we used the Qiskit Python library~\cite{Qiskit}.
We tested our approach on different molecules, for each molecule a series of 50 simulations were carried out.  We employed noiseless statevector simulation. As general settings for any VQE, we decided to use the Broyden–Fletcher–Goldfarb–Shanno (BFGS)~\cite{Flet87} algorithm with a convergence threshold set to $10^{-6}$, which corresponds to the tolerance on the gradient of the parameters. 
\section{Results}
\label{sec:res}Multi-QIDA builds a compact reference circuit starting from an approximated Quantum Mutual Information matrix, computed by quantum chemistry methods, iteratively adding and optimizing a layer defined on the desired range of correlation strength.
Here, we collect the results obtained from our Multi-QIDA approach compared to standard Hardware-Efficient Ansatze (HEA) with ladder topology. 
\subsection{Preprocessing: QMI matrices}
\label{ssec:qmi_matrix}
\begin{figure}[!htb]
    \centering
    \begin{subfigure}[b]{.460\linewidth}
    \includegraphics[width=\linewidth]{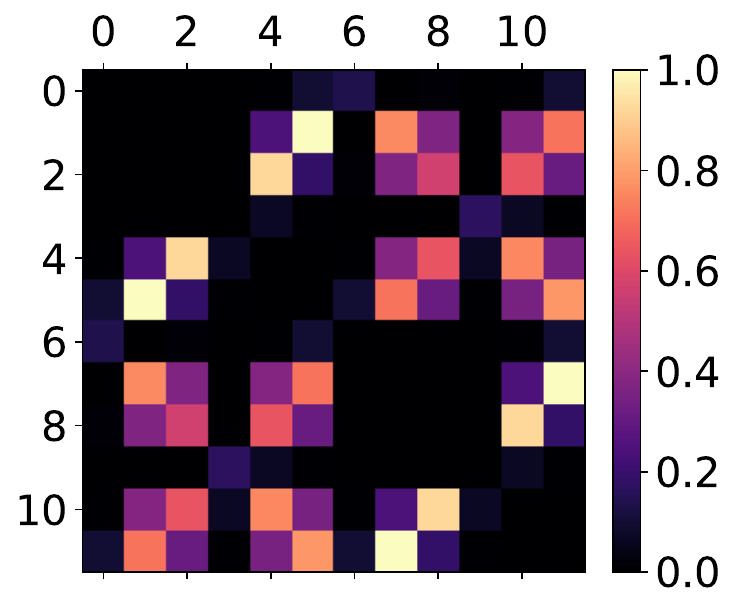}
    \caption{}\label{sfig:h2o_qmi}
    \end{subfigure}
    \begin{subfigure}[b]{.460\linewidth}
    \includegraphics[width=\linewidth]{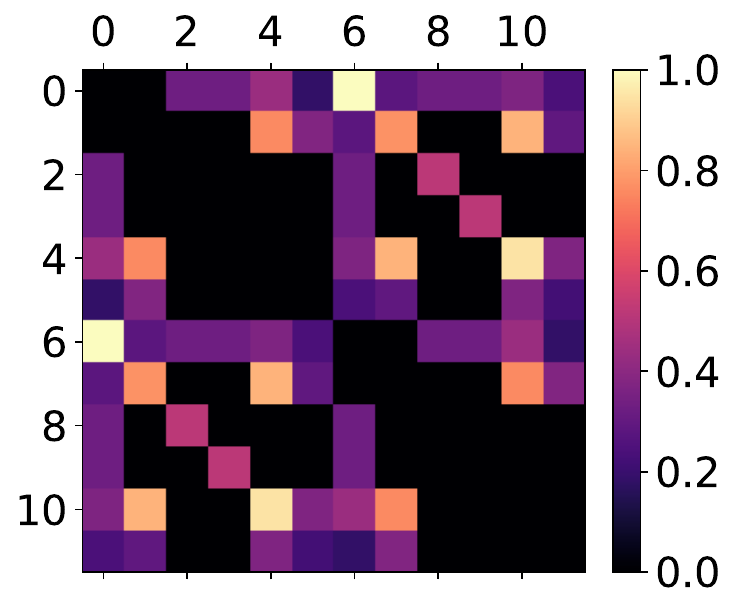}
    \caption{}\label{sfig:beh2_qmi}
    \end{subfigure}
    \begin{subfigure}[b]{.460\linewidth}
    \includegraphics[width=\linewidth]{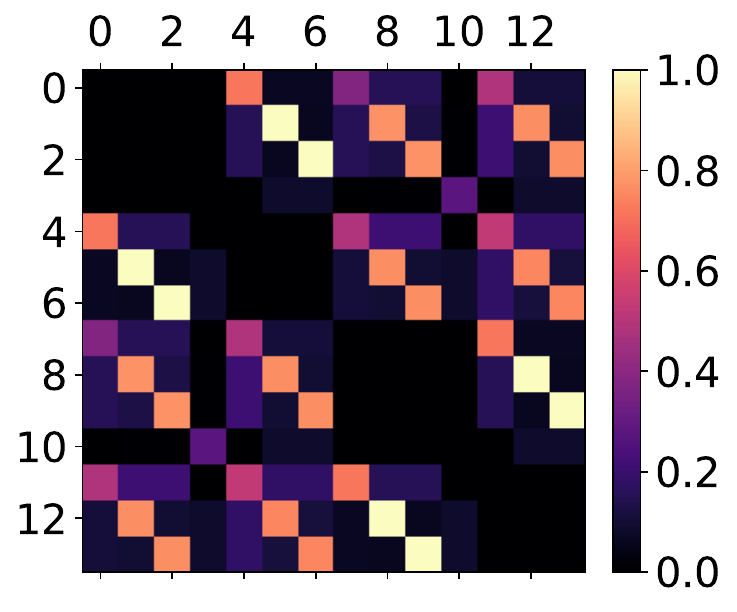}
    \caption{}\label{sfig:nh3_qmi}
    \end{subfigure}
    \begin{subfigure}[b]{.460\linewidth}
    \includegraphics[width=\linewidth]{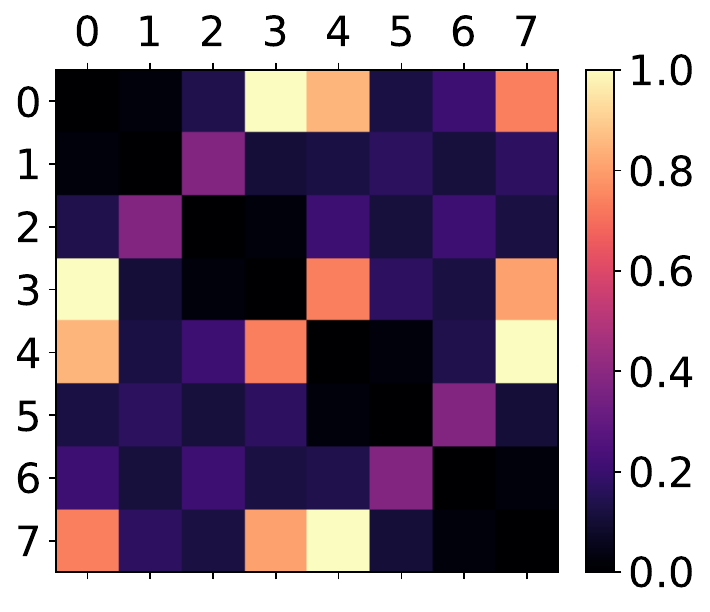}
    \caption{}\label{h2o_cas_qmi}
    \end{subfigure}
    \begin{subfigure}[b]{.50\linewidth}
    \includegraphics[width=\linewidth]{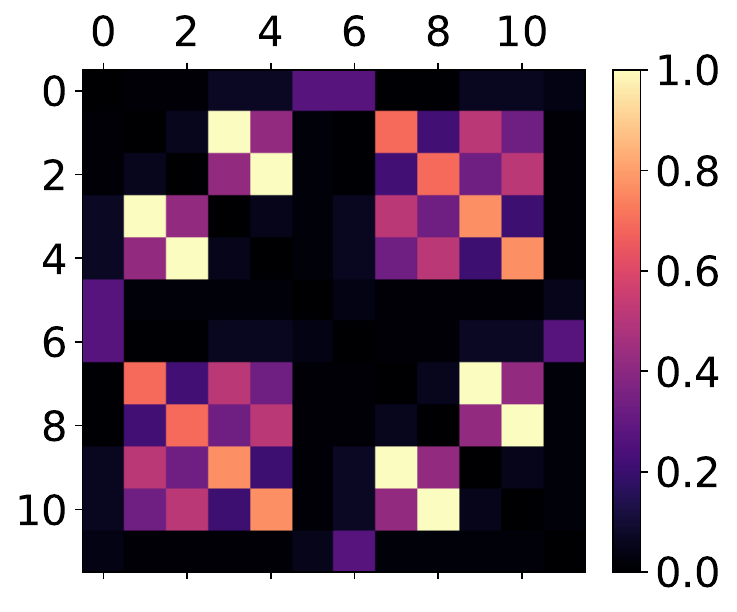}
    \caption{}\label{n2_cas_qmi}
    \end{subfigure}
    \caption{QMI matrices obtained by the tested systems: \textbf{(a)} H$_2$O INOs/12 qubits. \textbf{(b)} BeH$_2$ INOs/12 qubits. \textbf{(c)} NH$_3$ INOs/14 qubits. \textbf{(d)} H$_2$O CAS(4,4). \textbf{(d)} $N_2$ CAS(6,6). All obtained with SparQ at R-CISD level.}
\label{fig:fig_QMIs}
\end{figure}

Exploiting the SparQ algorithm, we computed the QMI matrices for all the systems listed in Table \ref{tab:mol_system}. For all the systems, an RCISD wavefunction has been used as a reference to build the QMI map. In particular, starting from the same set of orbitals used in the circuit, we used the PySCF RCISD solver in order to obtain the relevant information used by SparQ to build the QMI matrix. The settings used for SparQ are a cutoff of $10^{-12}$ for the Slater Determinant (SD) coefficients, and a maximum of $10^5$ SDs to build the approximated wavefunction. Then, for each system, by observing the distribution of the mutual-information pairs, the selection of the finesse-ratios is performed. Starting from the QMI matrices, shown in Figure \ref{fig:fig_QMIs}, the different finesse-ratios used are:
\begin{itemize}
    \item H$_2$O INOs: [0.5, 0.3, 0.1]
    \item BeH$_2$ INOs: [0.7, 0.4, 0.35, 0.3, 0.2]
    \item NH$_3$ INOs: [0.75, 0.5, 0.25, 0.2]
    \item H$_2$O 6-31G CAS(4,4): [0.5, 0.20, 0.15]
    \item N$_2$ cc-pVTZ CAS(6,6): [0.80, 0.6, 0.4, 0.2]
\end{itemize}
The choice of the finesse-ratio is done accordingly to the criteria defined in Section \ref{ssec:selection_crit}, in particular, we avoid the creation of a highly populated candidate set, we cover all the qubits with at least one QIDA-layer, and we stop after reaching a layer below 0.2 of QMI value.
The number of QIDA-layers is different for each system, but generally, we need to encode at least 3 layers in order to recover highly correlated pairs, mid-correlation, and low-lying correlations. As in the previous work, we completed the series of QIDA-layers with an additional ladder in order to join together disjointed groups of qubits.
\subsection{Performance analysis}
\label{ssec:performance}

In this section, the energetic comparison between HEA and Multi-QIDA, with also the deviation from the best performing VQE of both percentage correlation energy and absolute energy, are shown.

The results presented in Table \ref{tab:ino_systems_energies_main} and Table \ref{tab:cas_systems_energies_main} demonstrate that the proposed Multi-QIDA approaches, $QIDA_{\text{max}}$ (reduction of each QIDA-layer using a MST that maximizes the total QMI value) and $QIDA_{\text{emp}}$(reduction of each QIDA-layer using a mST that minimizes the distance between each qubit),  consistently outperform the standard ladder ansatz $(L)_{\text{CX}}^d$ in terms of correlation energy and absolute energy across all tested molecular systems. 
\begin{table}[!htb]
\centering
\fontsize{8pt}{8pt}
    \centering
    \setlength\tabcolsep{2pt}
   \begin{tabular}{|l|ccc|ccc|ccc|}\hline
 & \multicolumn{3}{c|}{BeH$_2$} \\ \hline
 & $(L)_6^{cx}$ & $QIDA_{max}$ & $QIDA_{emp}$  \\ \hline
$\epsilon_{avg}[\%]$ & $21.25(14.55)$ & \textbf{79.78}$(10.22)$ & \textbf{80.46}$(9.54)$ \\
$E_{avg}[Ha]$ & -$3.9146$ & $-3.93490$ & $-3.93510$  \\
$\epsilon_{best}[\%]$ & $49.75$ & \textbf{90.9} & \textbf{84.03} \\
$E_{best}[Ha]$ & -$3.92441$ & -$3.93872$ & -$3.93637$ \\ \hline
$M\epsilon D_{best}[\%]$ & $28.17$ & $10.98$ & $3.57$  \\
$MED_{best}[Ha]$ & $0.00980$ & $0.00380$ & $0.00120$\\
\hline\hline
 &\multicolumn{3}{c|}{H$_2$O} \\ \hline
 &$(L)_5^{cx}$ & $QIDA_{max}$ & $QIDA_{emp}$ \\ \hline
$\epsilon_{avg}[\%]$& -$111.50(284.70)$ & \textbf{89.52}$(2.31)$ & \textbf{89.71}$(2.31)$ \\
$E_{avg}[Ha]$& -$23.45440$ & $-23.55380$ & $-23.55390$ \\
$\epsilon_{best}[\%]$ & $81.56$ & \textbf{91.51} & \textbf{91.55}\\
$E_{best}[Ha]$ & -$23.54985$ & -$23.55477$ & -$23.55479$ \\ \hline
$M\epsilon D_{best}[\%]$ & $193.07$ & $1.99$ & $1.85$  \\
$MED_{best}[Ha]$ & $0.09550$ & $0.00100$ & $0.00090$  \\\hline\hline
 &  \multicolumn{3}{c|}{NH$_3$} \\ \hline
 & $(L)_5^{cx}$ & $QIDA_{max}$ & $QIDA_{emp}$ \\ \hline
$\epsilon_{avg}[\%]$& -$71.13(248.89)$ & \textbf{54.12}$(0.36)$ & \textbf{54.10}$(0.36)$ \\
$E_{avg}[Ha]$ & -$20.00260$ & $-20.08510$ & $-20.08510$ \\
$\epsilon_{best}[\%]$ & $41.22$ & \textbf{54.95} & \textbf{54.95} \\
$E_{best}[Ha]$ & -$20.07658$ & -$20.08561$ & -$20.08561$ \\ \hline
$M\epsilon D_{best}[\%]$ & $112.36$ & $0.83$ & $0.85$ \\
$MED_{best}[Ha]$ & $0.07400$ & $0.00050$ & $0.00060$ \\
\hline
\end{tabular}
\caption{BeH$_2$,H$_2$O, and NH$_3$ INOs system results.}
\label{tab:ino_systems_energies_main}
\end{table}
\begin{itemize}
    \item\textit{INOs systems: } In the simulations of BeH$_2$ , H$_2$O , and NH$_3$, Multi-QIDA circuits consistently achieve higher average percentage correlation energy ($\epsilon_{\text{avg}} \%$). In particular, for BeH$_2$, $QIDA_{\text{max}}$ achieves an $\epsilon_{\text{avg}}$ of $79.78\%$, a significant improvement over the $21.25\%$ obtained using the ladder ansatz, with a lower standard deviation as well ($10.22\%$ versus $14.55\%$). Better results are obtained by $QIDA_{emp}$, which increases the $\epsilon_{avg}$ up to $80.46\%$ 
Similarly, for H$_2$O, $QIDA_{\text{max}}$ obtains $89.52\%$, $QIDA_{emp}$ reaches a close $89.71\%$, whose compared to the negative correlation energy ($-111.50\%$) from $(L)_{\text{CX}}^5$, indicating that the ladder ansatz optimization is failing to converge, leading to a state with energy higher than HF. In this case, the deviation is two orders of magnitude lower for the $QIDA$ circuit. For NH$_3$, while all methods show similar results, $QIDA_{\text{max}}$ and $QIDA_{\text{emp}}$ marginally outperform the ladder topology with $\epsilon_{\text{avg}}$ values of $-54.12\%$ and $54.10\%$, respectively, compared to the negative value of $-71.13\%$ for $(L)_{\text{CX}}^d$. Here, both $QIDA_{max}$ and $QIDA_{emp}$ obtain close values of percentage correlation energy deviation, close to zero. In terms of best VQE results, BeH$_2$, $QIDA$ obtains a clear increase of $40\%$ over the ladders, whereas for the other systems, the increase is around $10/12\%$. In particular, for BeH$_2$,  Multi-QIDA circuits do not fall far from the average case, as expected. Compared to the $49.75\%$ obtained by the ladder, $QIDA_{max}$ reaches $90.90\%$, while $QIDA_{emp}$ gets $84.03\%$. For the correlation energy of best performing VQE of H$_2$O, we can observe that the HEA reaches a good $81.56\%$, while $QIDA_{max}$ and $QIDA_{emp}$ get similar values, around $91\%$. Finally, for NH$_3$, the ladder reaches $41.22\%$ correlation energy, while Multi-QIDA in both settings reaches $54.94\%$.
    In terms of percentage correlation energy deviation, $QIDA$ shows a lower dispersion w.r.t. the best-performing VQE, hitting a dispersion of two orders of magnitude lower than ladder ansatz for H$_2$O and NH$_3$.
\end{itemize}
In general, we can find a consistent number of VQEs for the ladder topology, which due to the random initial parametrization, are guided in a completely wrong energetic solution, way lower than the HF energy.
\begin{table}[!htb]
\centering
\fontsize{8pt}{8pt}
    \centering
    \setlength\tabcolsep{2pt}
   \begin{tabular}{|l|ccc|ccc|}\hline
 & \multicolumn{3}{c|}{H$_2$O $CAS(4,4)$}\\ \hline
 & $(L)_5^{cx}$ & $QIDA_{max}$ & $QIDA_{emp}$  \\ \hline
$\epsilon_{avg}[\%]$ &$55.42(27.46)$ & \textbf{82.36}$(6.32)$ & \textbf{80.32}$(9.18)$\\ 
$E_{avg}[Ha]$&$-6.62720$ & $-6.62800$ & $-6.62790$  \\ 
$\epsilon_{best}[\%]$&$92.17$ & \textbf{95.42} & \textbf{97.81}  \\
$E_{best}[Ha]$&$-6.62828$ & $-6.62837$ & $-6.62844$  \\ \hline
$M\epsilon D_{best}[\%]$&$36.75$ & $13.06$ & $17.49$\\
$MED_{best}[Ha]$ &$0.00110$ & $0.00040$ & $0.00050$ \\\hline\hline
 & \multicolumn{3}{c|}{$N_2$ $CAS(6,6)$}\\ \hline
 &  $(L)_6^{cx}$ & $QIDA_{max}$ & $QIDA_{emp}$ \\ \hline
$\epsilon_{avg}[\%]$ & $20.85(79.43)$ & \textbf{57.13}$(6.53)$ & \textbf{58.48}$(17.59)$  \\ 
$E_{avg}[Ha]$& $-11.44410$ & $-11.46860$ & $-11.46950$ \\ 
$\epsilon_{best}[\%]$& $79.99$ & \textbf{82.17} & \textbf{85.22} \\
$E_{best}[Ha]$& $-11.48403$ & $-11.48551$ & $-11.48757$  \\ \hline
$M\epsilon D_{best}[\%]$ & $59.13$ & $25.04$ & $26.75$\\
$MED_{best}[Ha]$ & $0.04000$ & $0.01690$ & $0.01810$\\\hline 
\end{tabular}
\caption{H$_2$O 6-31G CAS(4,4) and $N2_2$ cc-PVTZ CAS(6,6) system results. }
\label{tab:cas_systems_energies_main}
\end{table}
\begin{itemize}
    \item \textit{Active Region systems}: The second group of simulations are related to the application of Multi-QIDA in more complex systems, which are considered by dividing orbitals into inactive and active space regions. For the H$_2$O CAS(4,4) system, $QIDA_{\text{max}}$ achieves an $\epsilon_{\text{avg}}$ of $82.36\%$ while $QIDA_{emp}$ reaches a $80.32\%$, which are both enhanced performance with respect to the HEA ladder, which obtains on average $55.42\%$ correlation energy. The same behavior can be found for the $N_2$ CAS(6,6), for which the standard ladder obtains on average $20.85\%$ of correlation energy, while $QIDA_{max}$ is able to reach $57.13\%$ and $QIDA_{emp}$, a slightly higher value of $58.48\%$. In terms of best-performing VQE results, we have that the results of HEA ladders and Multi-QIDA circuits are close, but in any case, the latter reaches slightly higher correlation energy. 

\end{itemize}

Comparing instead the results obtained in terms of percentage correlation energy deviation w.r.t. to the best-performing VQE, we can notice that $Multi-QIDA$ behaves clearly better for complete systems, so in our case INOs systems, while it has a higher dispersion for Active space systems, which may be related to the fact that we are not including any kind of double excitations directly in the ans\"atze. Also, we can observe that there is no clear distinction between the two type of selection performed on the QIDA-layers, a difference that may be appreciated more if applied in the context of real hardware topology or real devices.

\subsection{Convergence and Precision}
\begin{figure}[!htb]
    \centering
    \includegraphics[width=1.02\linewidth]{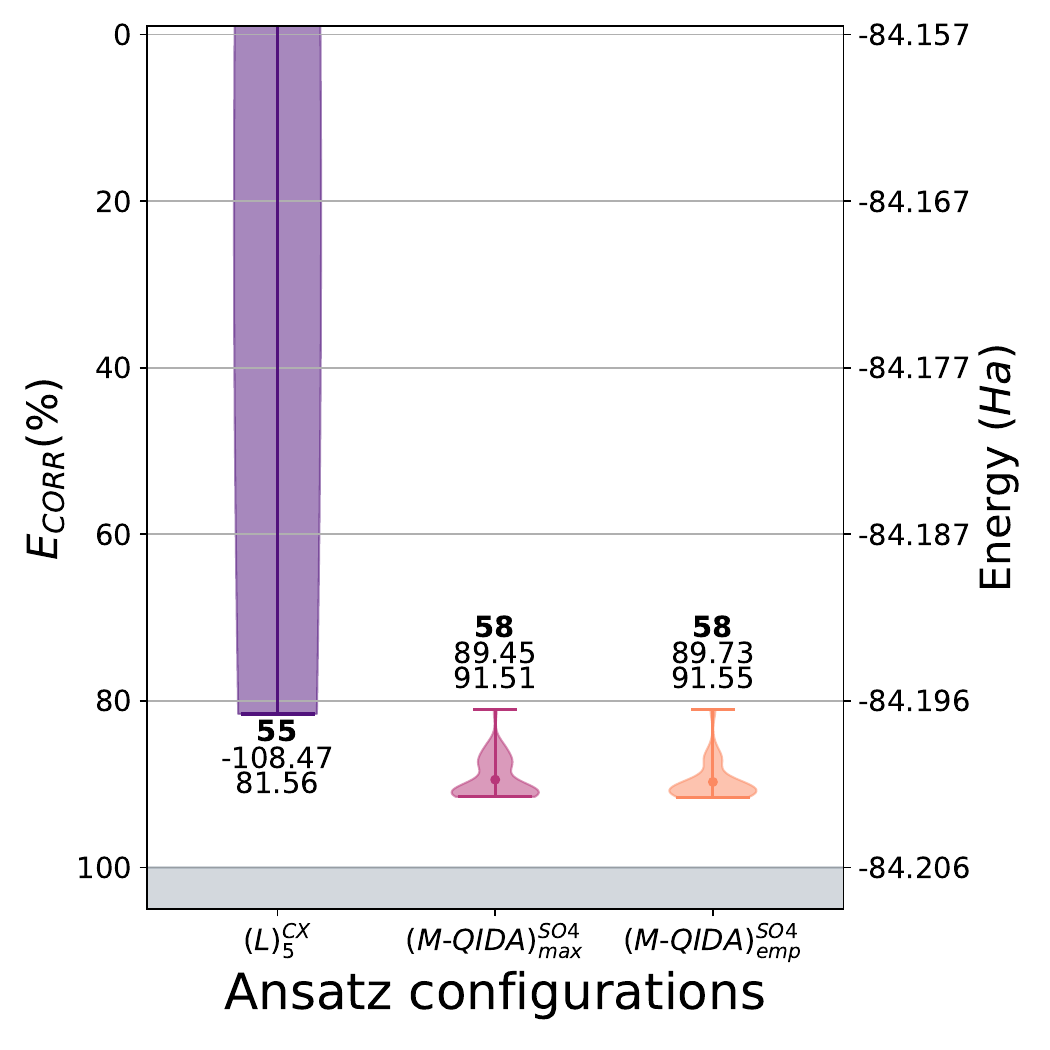}
\caption{ H$_2$O INOs system comparison between depth 5 ladder HEA and Multi-QIDA in both max and empirical configuration. The three numbers associated with each violin show the \textbf{number of CNOTs}, $\epsilon_{avg}$, and $\epsilon_{best}$, starting from the upper one, for each simulation setting.}
    \label{fig:h2o_nos_violin}
\end{figure}
Here, we briefly analyze the results presented in Figure \ref{fig:h2o_nos_violin} and in the Appendix Figures \ref{fig:beh2_nos_violin}-\ref{fig:n2_cas_violin}, related to the precision of VQE runs, and Figure \ref{fig:h2o_ino_conv} and in the Appendix Figures \ref{fig:beh2_ino_conv}-\ref{fig:n2_cas_conv}, related to the dispersion of the optimizations.
Each of the plots in the first group represents the energy, $E$, and the percentage correlation energy, $E_{corr}$, for every system and for all three ansatz configurations. In these plots, the results for each ansatz configuration are represented by a violin plot. The width of each violin is related to the frequency of the VQE outcomes. They are useful for assessing the consistency of the algorithm (how clustered or spread out the results are) and identifying trends, such as whether the algorithm reliably converges to a minimum energy or exhibits variability. At the two extremes of each violin are presented the worst and the best performing VQE results, while the central dot represents the average value.

We can notice that standard HEA with ladder-fashion connectivity presents results that are way lower than HF and thus the average value is strongly shifted from the best-performing VQE. In the systems in which this behaviour does not happen, the average result is anyway distant than the optimal one and this means that it is required to re-run the circuit even more time, compared to Heisenberg Model Hamiltonian, \cite{tarocco2024} before obtaining a satisfying result. Ladders quickly encounter and falls into local minima that are far away from the best  result. An example is in Figure \ref{fig:h2o_nos_violin}, for which HEA can reach a good 81,54\% of correlation energy but is heavily penalized on average because most of the runs falls below HF energy, while Multi-QIDA is able to maintain, as expected, a low dispersion around the best-performing VQE. We can further identify some difficulties of Multi-QIDA in correctly describing the NH$_3$ ground state and in particular, it get stuck around 50\% of correlation energy, shown in Figure \ref{fig:nh3_nos_violin}. But in general, we can see that even in the worst-performing VQE, for which the population is very small, the energy is still higher than the average case of HEA and in some cases higher that the best-performing ladder VQE. This last case is shown in Figure \ref{fig:h2o_nos_violin}. In Figure \ref{fig:n2_cas_violin}, we can see how, for the $N_2$ CAS(6,6) the performance of the two Multi-QIDA selection criteria is completely different, and in particular, the \textit{maximum correlation} fails to compact the results towards the best-performing VQE but on the worst-performing, still reaching higher correlation energy than the average HEA circuits. Another example is for BeH$_2$, in which the best-performing for ladders reaches $\sim$50\% correlation energy, while a very limited portion of VQEs for Multi-QIDA fails to follow the right variational path. Multi-QIDA approach, as already seen for spin systems, can guide the variational wavefunction in the right spot, in an iterative and adaptive way, without the requirement of building and optimizing the full variational space from the beginning.
\begin{figure}[!htb]
    \centering
    \includegraphics[scale=0.45]{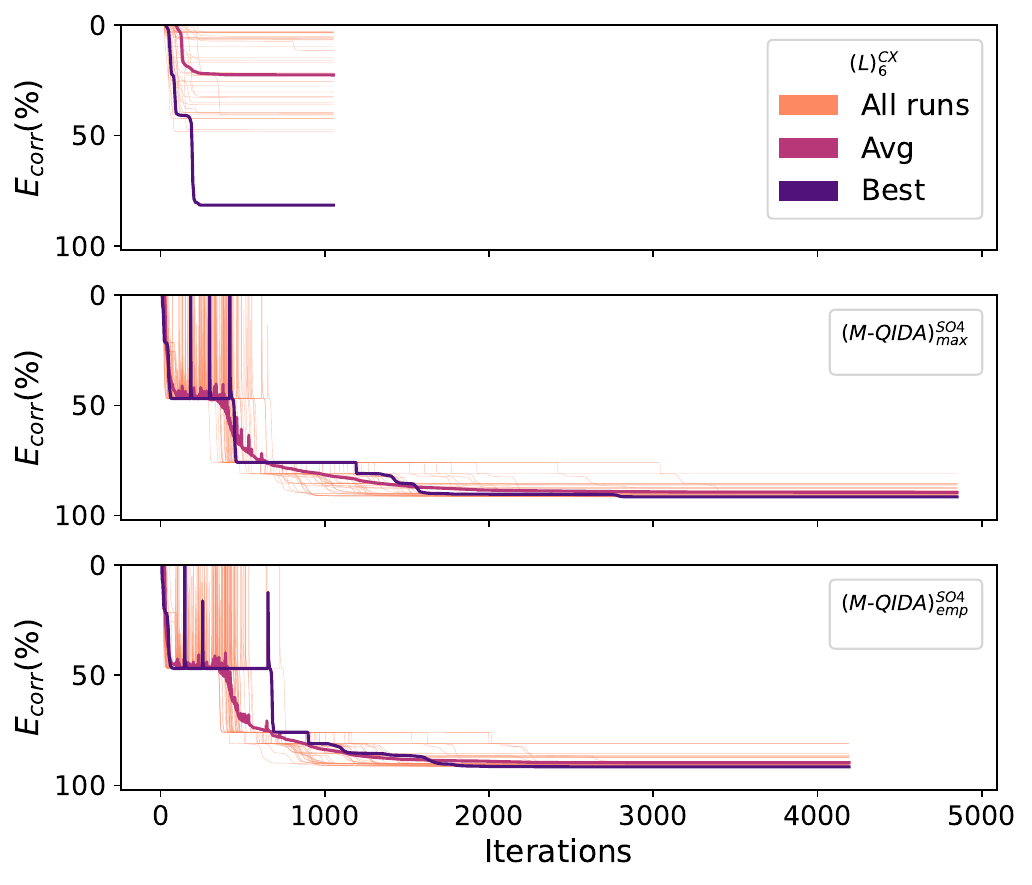}
    \caption{H$_2$O INOs system convergence trajectories for each of the 50 VQEs. In particular, starting from the upper plot $(L)^5$, Multi-QIDA with \textit{max} selection criteria, and last Multi-QIDA with \textit{emp} reduction.}
    \label{fig:h2o_ino_conv}
\end{figure}
This behavior can also be noticed in the second group of plot, Figure \ref{fig:h2o_ino_conv} and in Appendix Figures \ref{fig:beh2_ino_conv}-\ref{fig:n2_cas_conv}, in which we show the trajectories of the convergence of each VQE for all the systems. For HEA simulations, the upper subplot, it is clear that the population of VQE that are actually getting towards the right optimization path is very low compared to the one that diverges or gets stuck in local minima. For Multi-QIDA instead, it is possible to notice that the trajectories, even if they get perturbed at each additional layer, tend to be more compact and closer to the best-performing and average trajectory. From the best-performing VQE, which is the bolder trajectory, it is also possible to notice the quick recovery and restart of the optimization after being perturbed, allowing the escape from the previous local minima and without ending in a higher convergence point.

As in the previous work, we are aware of the higher computational cost required by Multi-QIDA to converge and to end the full optimization procedure. The average number of iteration is usually two/three times the number of iterations required by the corresponding HEA ansatz, and most of the optimization procedure is wasted in the relaxation procedure.

\subsection{Wavefunction Properties} 
\label{ssec:symmetries}
Together with the measurement of the performance, we decided to define also metrics to evaluate the capability of the Multi-QIDA ansatz to satisfy symmetry constraints and fidelity w.r.t. the exact ground state. 
Given the fact that all the system studied are closed shell, the $\hat{S_z}$ and $\hat{S}^2$ are both zero, while the number of particles, $\hat{N_e}$, for each specific INOs system is: BeH$_2$=4 , H$_2$O=6, and NH$_3$=8 , for CAS system instead: H$_2$O $CAS(4,4)$=4 and $N_2$ $CAS(6,6)$=6. The properties analysis results are collected in Table \ref{tab:ino_systems_props_main} for INOs systems, and in Table\ref{tab:cas_systems_props_main} for Active-Space systems.
\begin{table}[!htb]
\centering
\fontsize{9pt}{6pt}
    \centering
   \begin{tabular}{|l|ccc|ccc|ccc}\hline
 & \multicolumn{3}{c|}{BeH$_2$}\\ \hline
 & $(L)_6^{cx}$ & $QIDA_{max}$ & $QIDA_{emp}$ \\ \hline
$\mathcal{F}_{avg}[\%]$ & $98.89579 $ & \cellcolor[HTML]{C0FAB4}$ 99.60117 $ & \cellcolor[HTML]{C0FAB4}$99.61037 $  \\
$\hat{S_z}_{avg}$ & -$0.00002 $ & $ 0 $ & $0 $  \\
$\hat{S}^2_{avg}$ & $0.00261 $ & \cellcolor[HTML]{C0FAB4}$ 0.00111 $ & \cellcolor[HTML]{C0FAB4}$0.00062 $  \\
$\hat{N}_{avg}$ & $4.00001 $ & $ 4.00000 $ & $3.99999 $ \\ \hline
$\mathcal{F}_{best}[\%]$ & $99.28972 $ & \cellcolor[HTML]{C0FAB4}$ 99.80993 $ & \cellcolor[HTML]{C0FAB4}$99.65462 $ \\
$\hat{S_z}_{best}$ & $0 $ & $ 0 $ & $0.00001 $\\
$\hat{S}^2_{best}$ & $0.00016 $ & \cellcolor[HTML]{FFCCC9}$ 0.00026 $ & $0.00002 $ \\
$\hat{N}_{best}$ & $4.00000 $ & \cellcolor[HTML]{FFCCC9}$ 4.00010 $ & $3.99999 $ \\\hline\hline
 & \multicolumn{3}{c|}{H$_2$O} \\ \hline
& $(L)_5^{cx}$ & $QIDA_{max}$ & $QIDA_{emp}$ \\ \hline
$\mathcal{F}_{avg}[\%]$ & $89.08740 $ & \cellcolor[HTML]{C0FAB4}$99.86803 $ & \cellcolor[HTML]{C0FAB4}$99.86958 $ \\
$\hat{S_z}_{avg}$ & $0.10343 $ & \cellcolor[HTML]{C0FAB4}$0 $ & \cellcolor[HTML]{C0FAB4}$0 $\\
$\hat{S}^2_{avg}$ & $0.22269 $ & \cellcolor[HTML]{C0FAB4}$0.00128 $ & \cellcolor[HTML]{C0FAB4}$0.00115$\\
$\hat{N}_{avg}$ & $7.97648 $ & \cellcolor[HTML]{C0FAB4}$7.99998 $ & \cellcolor[HTML]{C0FAB4}$7.99998 $\\ \hline
$\mathcal{F}_{best}[\%]$ & $99.77667 $ & \cellcolor[HTML]{C0FAB4}$99.88329 $ & \cellcolor[HTML]{C0FAB4}$99.88125 $  \\
$\hat{S_z}_{best}$ & $0 $ & $0 $ & $0.00001 $\\
$\hat{S}^2_{best}$ & $0.00001 $ & \cellcolor[HTML]{FFCCC9}$0.00015 $ & \cellcolor[HTML]{FFCCC9}$0.00042 $\\
$\hat{N}_{best}$ & $8.00000 $ & $8.00000 $ & $7.99998 $ \\ \hline\hline

 &  \multicolumn{3}{c|}{NH$_3$} \\ \hline
 & $(L)_5^{cx}$ & $QIDA_{max}$ & $QIDA_{emp}$ \\ \hline
$\mathcal{F}_{avg}[\%]$ & $92.23381 $ & \cellcolor[HTML]{C0FAB4}$99.19447 $ & \cellcolor[HTML]{C0FAB4}$99.19421 $ \\
$\hat{S_z}_{avg}$& $0.05542 $ & \cellcolor[HTML]{C0FAB4}$0 $ & \cellcolor[HTML]{C0FAB4}$0 $ \\
$\hat{S}^2_{avg}$ & $0.13443 $ & \cellcolor[HTML]{C0FAB4}$0.00003 $ & \cellcolor[HTML]{C0FAB4}$0.00004 $ \\
$\hat{N}_{avg}$  & $7.98916 $ & \cellcolor[HTML]{C0FAB4}$8.00000 $ & \cellcolor[HTML]{C0FAB4}$8.00000 $ \\ \hline
$\mathcal{F}_{best}[\%]$ & $98.98914 $ & \cellcolor[HTML]{C0FAB4}$99.20485 $ & \cellcolor[HTML]{C0FAB4}$99.20485 $ \\
$\hat{S_z}_{best}$ & $0 $ & $0 $ & $0 $ \\
$\hat{S}^2_{best}$ & $0.00002 $ & $0 $ & $0 $ \\
$\hat{N}_{best}$ & $7.99998 $ & $8.00000 $ & $8.00000 $ \\ \hline
%$M\mathcal{F}D_{best}$ & $0.06760 $ & \cellcolor[HTML]{C0FAB4}$0.00010 $ & \cellcolor[HTML]{C0FAB4}$0.00010 $ \\
%$M\hat{S}_zD_{best}$ & $0.05540 $ & \cellcolor[HTML]{C0FAB4}$0 $ & \cellcolor[HTML]{C0FAB4}$0 $ \\
%$M\hat{S}^2D_{best}$ & $0.13440 $ & \cellcolor[HTML]{C0FAB4}$0 $ & \cellcolor[HTML]{C0FAB4}$0 $ \\
%$M\hat{N}D_{best}$ & $0.01085 $ & \cellcolor[HTML]{C0FAB4}$0.00001 $ & \cellcolor[HTML]{C0FAB4}$0.00001 $
\end{tabular}
\caption{Properties of BeH$_2$, H$_2$O, and NH$_3$ INOs system.}
\label{tab:ino_systems_props_main}
\end{table}
\begin{itemize}
    \item \textit{INOs Systems :} On average, Multi-QIDA obtained an improvement on each of the property measured. For the Fidelity $\mathcal{F}$ with respect to the ground state, both the Multi-QIDA configuration recovered a slightly higher value, $\sim 1.30\%$ ,for BeH$_2$, a relevant increment of $\sim10.8\%$ for H$_2$O, and a non-negligible $\sim6.8\%$ for NH$_3$. For the spin symmetries, in both cases, Multi-QIDA has been able to improve the results or in general to not lower the quality of the result. In particular, for $\hat{S_z}$,  Multi-QIDA obtain 0 in the two selection criteria, while the HEA gets $0.10343$ for H$_2$O and $0.05541$. In terms of $\hat{S}^2$ instead, the improvements have been obtained on all the three systems, in particular, for BeH$_2$, Multi-QIDA with \textit{max} selection halved to 0.00111 the HEA value, 0.00261, while Multi-QIDA with \textit{emp} reduce the value lower to 1e-3. For H$_2$O, the value for HEA is 0.22268 and Multi-QIDA configuration reduced it of two orders of magnitude. The best improvement have been obtained for NH$_3$ for which the value obtained by HEA, 0.13443, has been reduced by four orders of magnitude. For the number of particles, $\hat{N}$, the values have been refined to exact values only for NH$_3$. In terms of best-performing VQE, as we expected, also standard HEA is able to recover almost the same values of properties as Multi-QIDA.
    \item \textit{Active Region systems :} For this systems, the main improvement of Multi-QIDA with respect to HEA can be found mainly for the second system, $N_2$ CAS(6,6). On average, Multi-QIDA obtains an slightly higher fidelity compared to HEA and for $N_2$ closer $\hat{S}^2$ and $\hat{N}$ to the exact value. As before and as we expected, the best-performing VQE of HEA is able to obtain properties values closer to the one measure from a Multi-QIDA circuit.
    \begin{table}[!htb]
\centering
\fontsize{9pt}{6pt}
    \centering
   \begin{tabular}{|l|ccc|ccc|}\hline
 & \multicolumn{3}{c|}{H$_2$O $CAS(4,4)$} \\ \hline
 & $(L)_5^{cx}$ & $QIDA_{max}$ & $QIDA_{emp}$\\ \hline
$\mathcal{F}_{avg}[\%]$ & $99,95724 $ & \cellcolor[HTML]{C0FAB4}$ 99,98186$ & \cellcolor[HTML]{C0FAB4}$ 99,97978 $ \\
$\hat{S_z}_{avg}$ & $0 $ & $ 0 $ & $ 0 $  \\
$\hat{S}^2_{avg}$ & $0,00025 $ & $ 0,00035 $ & $ 0,00030 $  \\
$\hat{N}_{avg}$ & $4,00000 $ & $ 4,00000 $ & $ 4,00000 $ \\ \hline
$\mathcal{F}_{best}[\%]$ & $99,99057 $ & \cellcolor[HTML]{C0FAB4}$ 99,99390$ & \cellcolor[HTML]{C0FAB4}$ 99,9967 $ \\
$\hat{S_z}_{best}$ & $0$ & $ 0 $ & $ 0 $ \\
$\hat{S}^2_{best}$ & $0,00018$ & $ 0,00010 $ & \cellcolor[HTML]{C0FAB4}$ 0,00004 $\\
$\hat{N}_{best}$ & $4,00000$ & $ 4,00000 $ & $ 4,00000 $ \\ \hline\hline
 &  \multicolumn{3}{c|}{$N_2$ $CAS(6,6)$} \\ \hline
 & $(L)_5^{cx}$ & $QIDA_{max}$ & $QIDA_{emp}$ \\ \hline
$\mathcal{F}_{avg}[\%]$& $ 94,16438 $ & \cellcolor[HTML]{C0FAB4}$ 98,77033 $ & \cellcolor[HTML]{C0FAB4}$ 98,70980 $ \\
$\hat{S_z}_{avg}$ & $ -0,00007 $ & $ 0 $ & $ 0 $ \\
$\hat{S}^2_{avg}$ & $ 0,05629 $ & \cellcolor[HTML]{C0FAB4}$ 0,00049 $ & $ 0,04372 $ \\
$\hat{N}_{avg}$  & $ 6,04056 $ & \cellcolor[HTML]{C0FAB4}$ 6,00000 $ & \cellcolor[HTML]{C0FAB4}$ 6,00000 $ \\ \hline
$\mathcal{F}_{best}[\%]$ & $ 99,5714 $ & \cellcolor[HTML]{FFCCC9}$ 99,55640 $ & \cellcolor[HTML]{C0FAB4}$ 99,67421 $ \\
$\hat{S_z}_{best}$ & $ -0,0002 $ & $ 0 $ & $0 $ \\
$\hat{S}^2_{best}$ & $ 0,00192 $ & \cellcolor[HTML]{FFCCC9}$ 0,00766 $ & \cellcolor[HTML]{C0FAB4}$ 0 $ \\
$\hat{N}_{best}$  & $ 5,99956 $ & \cellcolor[HTML]{C0FAB4}$ 6,00000 $ & \cellcolor[HTML]{C0FAB4}$ 6,00000 $ \\ \hline
%$M\mathcal{F}D_{best}$ & $ 0,05410 $ & \cellcolor[HTML]{C0FAB4}$ 0,00790 $ & \cellcolor[HTML]{C0FAB4}$0,00960 $ \\
%$M\hat{S}_zD_{best}$ & $ 0,00020 $ & \cellcolor[HTML]{C0FAB4}$ 0 $ & \cellcolor[HTML]{C0FAB4}$0 $ \\
%$M\hat{S}^2D_{best}$ & $ 0,05440 $ & \cellcolor[HTML]{C0FAB4}$ 0,00790 $ & $0,04370 $ \\
%$M\hat{N}D_{best}$ & $ 0,04103 $ & \cellcolor[HTML]{C0FAB4}$ 0 $ & \cellcolor[HTML]{C0FAB4}$0 $
\end{tabular}
\caption{Properties of H$_2$O 6-31G CAS(4,4) and $N_2$ cc-PVTZ CAS(6,6) system.}
\label{tab:cas_systems_props_main}
\end{table}

\end{itemize}

\section{Discussion and Conclusions}
The Molecular System-Multi-QIDA method, an extension of the Quantum Information Driven Ansatz (QIDA) is designed to leverage quantum mutual information (QMI) to construct compact, shallow quantum circuits that are tailored on the main and most relevant correlations present in the system. The method generates an initial wavefunction using QMI-derived correlations and incrementally adds layers to recover missing correlations, compared to the stand-alone QIDA method. Combined with the including of SO(4) correlators and spanning-tree based reduction of the qubit-pairs, Multi-QIDA is able to achieve a balance between computational efficiency and circuit expressiveness, while guaranteeing reliable results in terms of good approximated wavefunctions.

This approach has been benchmarked across various molecular systems, including H2O, BeH2, NH3, and active-space models like N2 CAS(6,6) and H2O CAS(4,4). It consistently outperforms hardware-efficient ansatz (HEA) configurations in energy accuracy, while maintaining fidelity to the true ground state and respecting essential physical symmetries. Furthermore, by incorporating iterative optimization, Multi-QIDA mitigates the challenges of barren plateaus, offering scalability and convergence advantages. These features make it a strong candidate for use as a starting guess in more complex ansätze like ADAPT-VQE or sampling procedures,  such as Quantum Selected CI (QSCI).

In conclusion, the Multi-QIDA method demonstrates significant promise in constructing resource-efficient and accurate quantum circuits for molecular simulations. However, several open questions remain. Can the method maintain its performance and scalability as the size and complexity of molecular systems is increasad. How effectively can Multi-QIDA integrate with adaptive approaches like ADAPT-VQE or sampling methods to tackle strongly correlated systems while avoiding optimization bottlenecks. What modifications would be required for Multi-QIDA to perform robustly on real-world quantum devices subject to noise and decoherence. Can be extended including different correlators, such as single/double-qubit-based excitations or Givens rotations. Lastly, is Multi-QIDA generalizable to problem in which a correlation matrix can be defined amongst the elements of the system. Addressing these questions will further clarify the potential of Multi-QIDA in advancing quantum chemistry simulations.

In addition, at variance with respect to other empirical ansatz, the \textit{SO4} gates provides a notable advantage not only in term of variational energies but also in term of preserving wavefunction properties and symmetries, such as total spin, spin projection, and particle number, thus ensuring better capacities to describe molecular systems.
 
\section*{Acknowledgments}
The authors acknowledge funding from the European Union - Next Generation EU, Mission 4 - Component 1 - Investment 4.1 (CUP E11I22000150001).
The authors acknowledge funding from the MoQS program, founded by the European Union’s Horizon 2020 research and innovation under Marie Sklodowska-Curie grant agreement number 955479.
The authors acknowledge funding from Ministero dell’Istruzione dell’Università e della Ricerca (PON R \& I 2014-2020).
The authors also acknowledge funding from National Centre for HPC. Big Data and Quantum Computing - PNRR Project, funded by the European Union - Next Generation EU.\\
L.G. acknowledges funding from the Ministero dell'Università e della Ricerca (MUR) under the Project PRIN 2022 number 2022W9W423 through the European Union Next Generation EU.
\printbibliography

\appendix
\counterwithin{figure}{section}
\renewcommand{\thefigure}{\thesection\arabic{figure}}
\counterwithin{table}{section}
\renewcommand{\thetable}{\thesection\arabic{table}}
\begin{appendices}
\section{Complete results}
\newpage
\subsection{Violin plots and Trajectories}
Convergence percentage correlation energy/absolute energy HEA against Multi-QIDA configurations for INOs molecular systems.
\begin{figure}[!htb]
    \centering
    \includegraphics[width=0.9\linewidth]{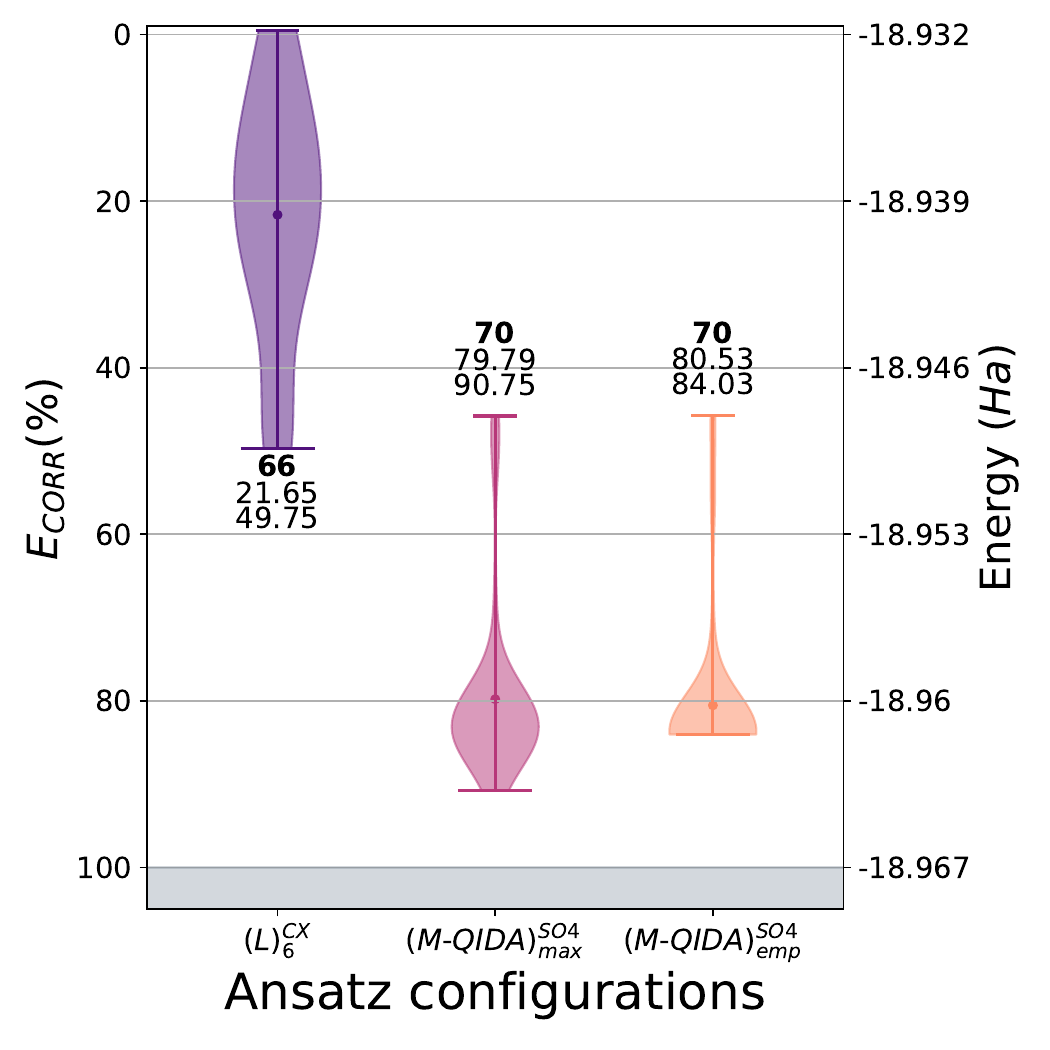}
    \caption{BeH$_2$ INOs system.}
    \label{fig:beh2_nos_violin}
\end{figure}
\begin{figure}[!htb]
    \centering
    \includegraphics[width=0.9\linewidth]{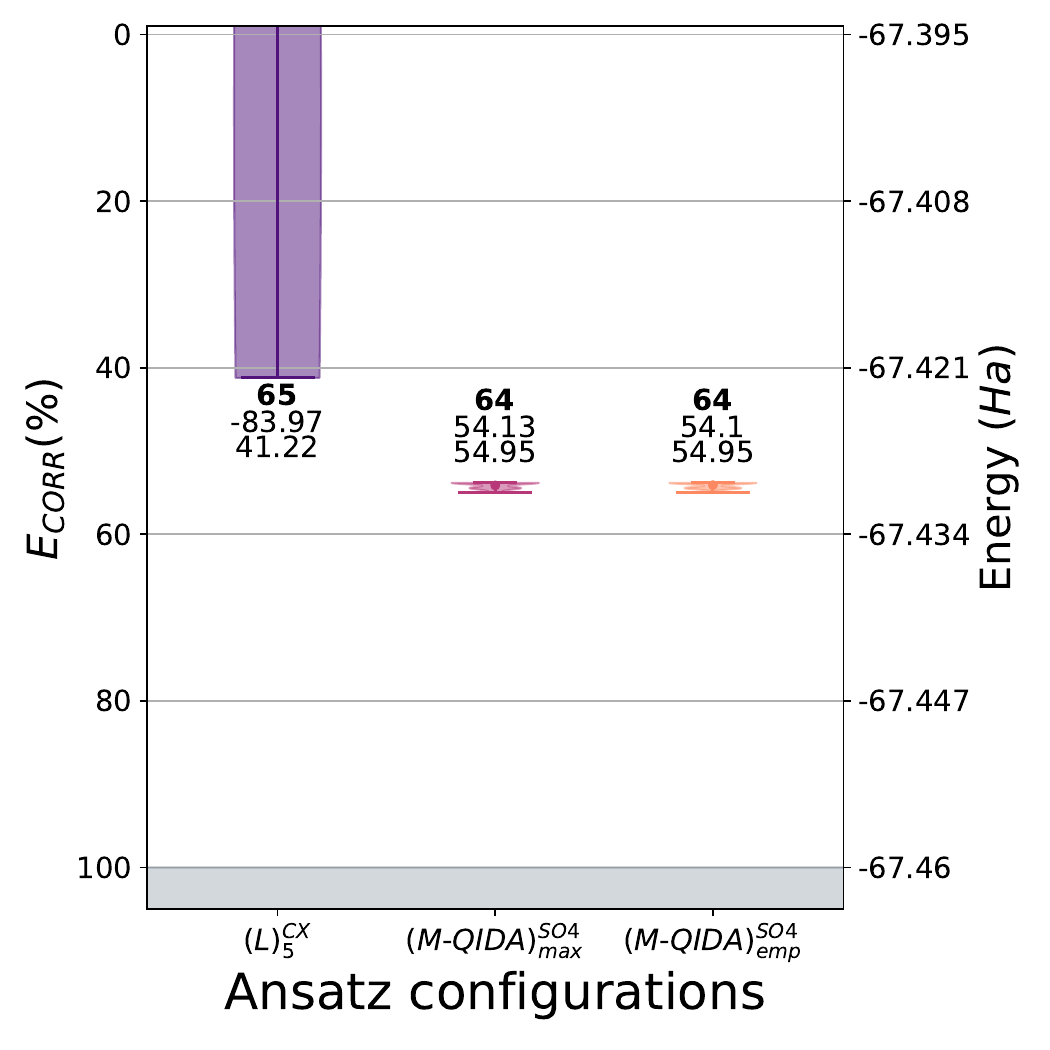}
    \caption{NH$_3$ INOs system.}
    \label{fig:nh3_nos_violin}
\end{figure}

Convergence percentage correlation energy/absolute energy HEA against Multi-QIDA configurations for CASCI/Active region molecular systems.
\begin{figure}[!htb]
    \centering
    \includegraphics[width=0.9\linewidth]{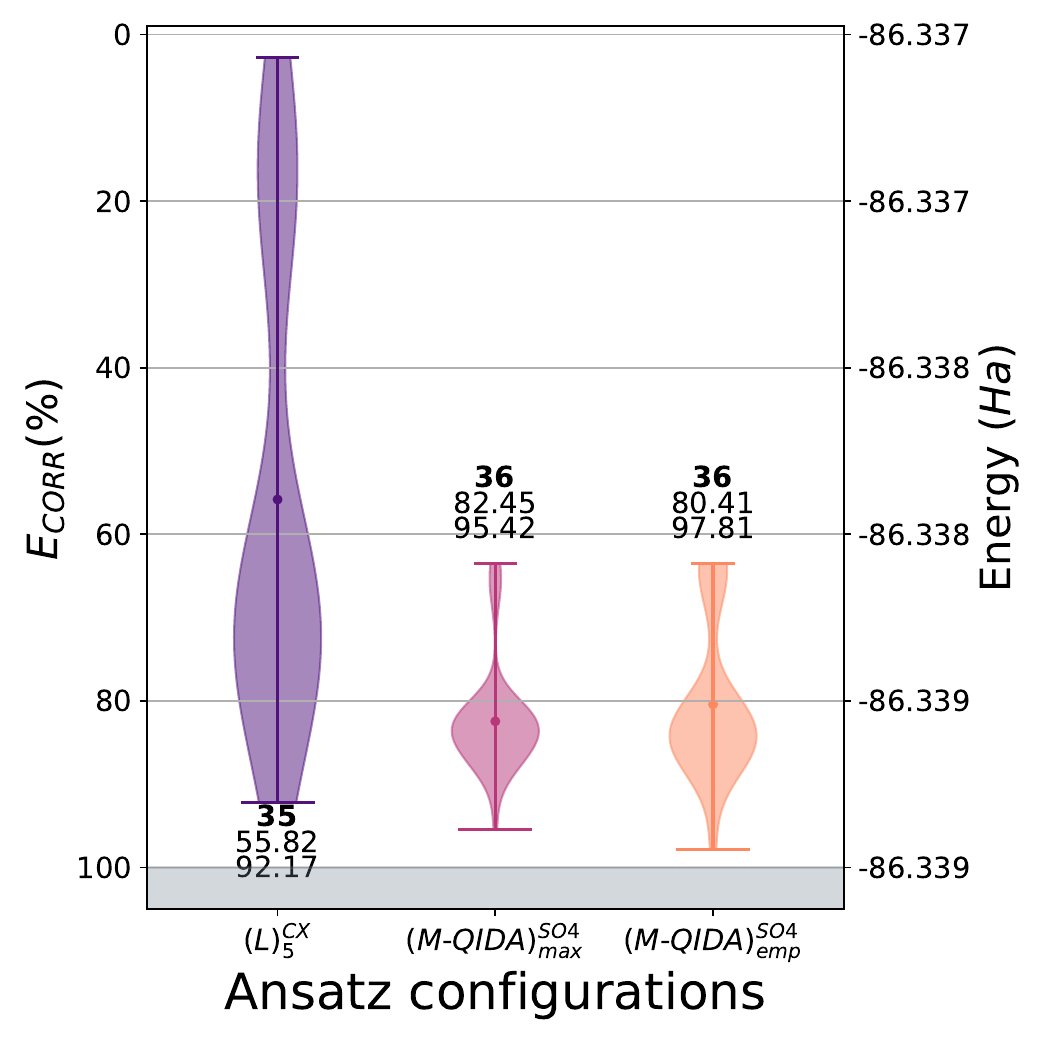}
    \caption{H$_2$O 6-31G CAS(4,4) system.}
    \label{fig:h2o_cas_violin}
\end{figure}
\begin{figure}[!htb]
    \centering
    \includegraphics[width=0.9\linewidth]{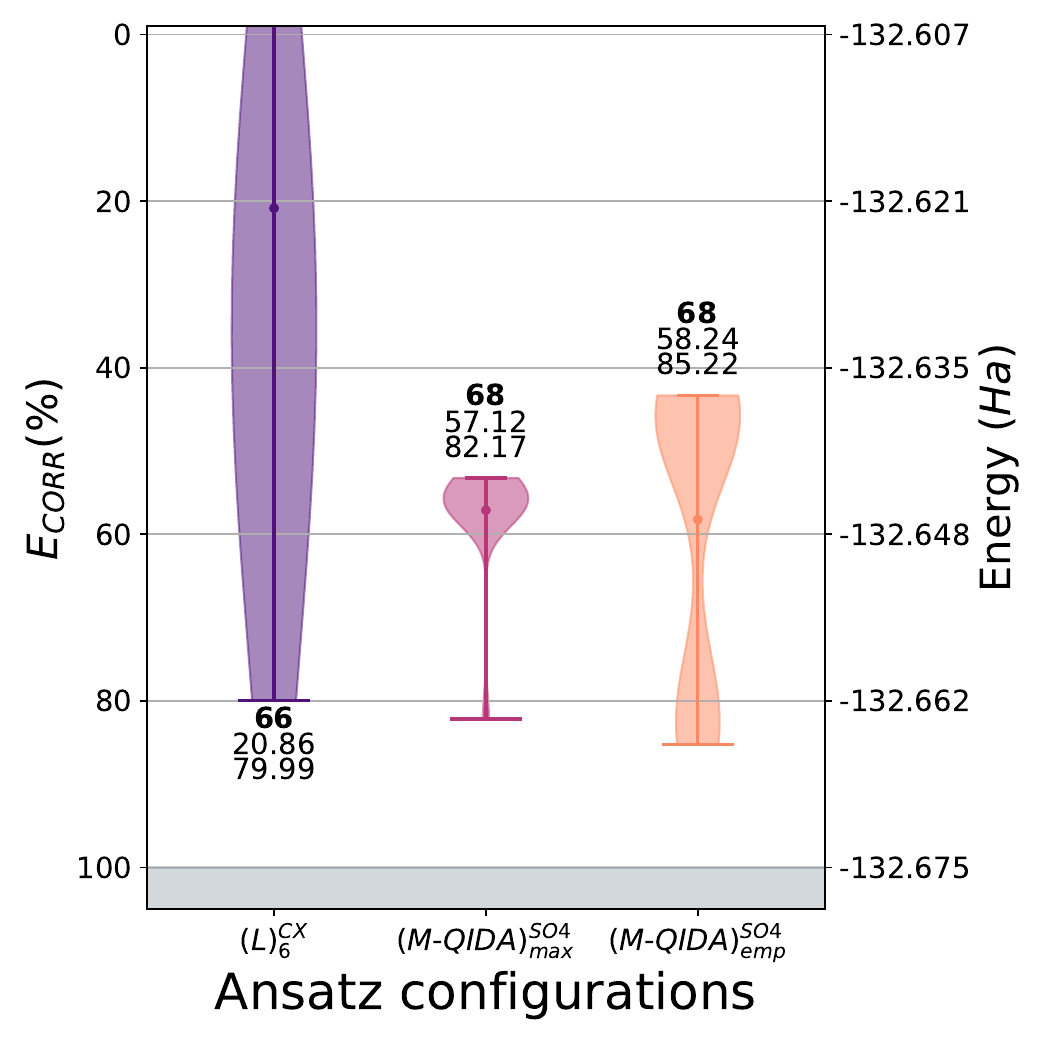}
    \caption{$N_2$ cc-PVTZ CAS(6,6) system.}
    \label{fig:n2_cas_violin}
\end{figure}

Optimization trajectories for each of the INOs system of 50 VQE for HEA ladder-fashion circuit against Multi-QIDA circuits. 

\begin{figure}[!htb]
    \centering
    \includegraphics[width=7cm, height=5.5cm]{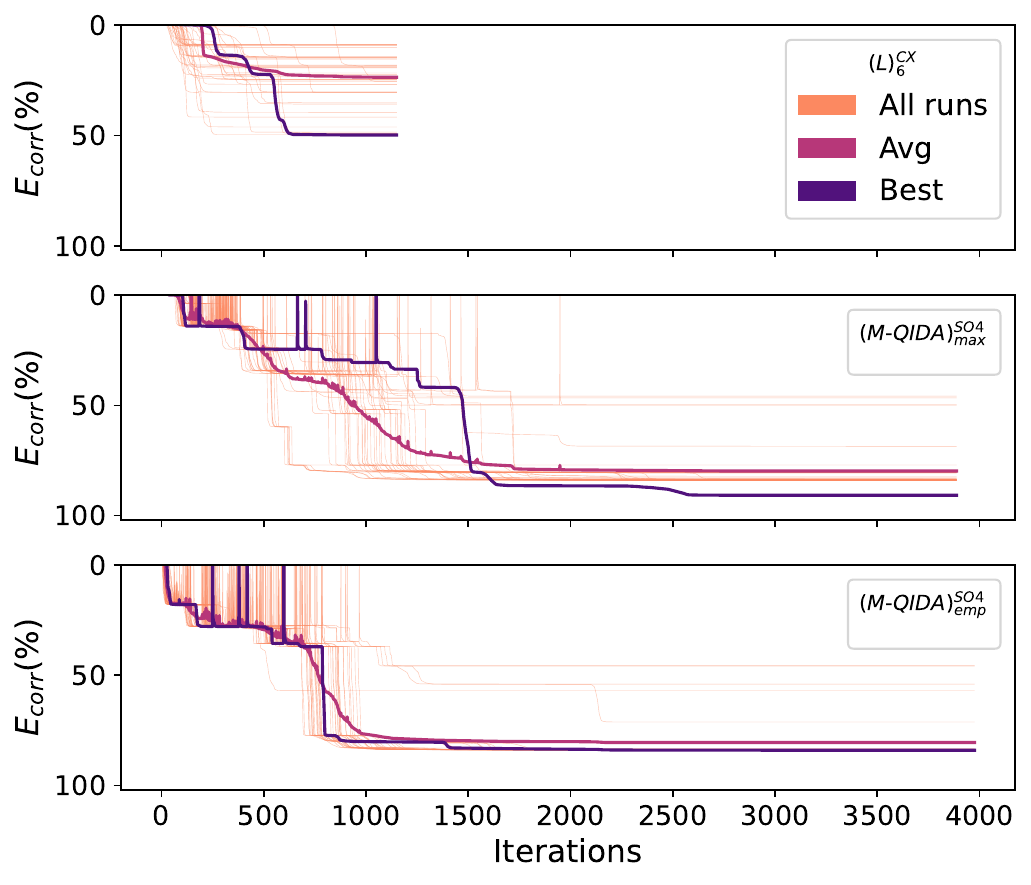}
    \caption{BeH$_2$ INOs system convergence trajectories}
    \label{fig:beh2_ino_conv}
\end{figure}

\begin{figure}[!htb]
    \centering
    \includegraphics[width=7cm, height=5.5cm]{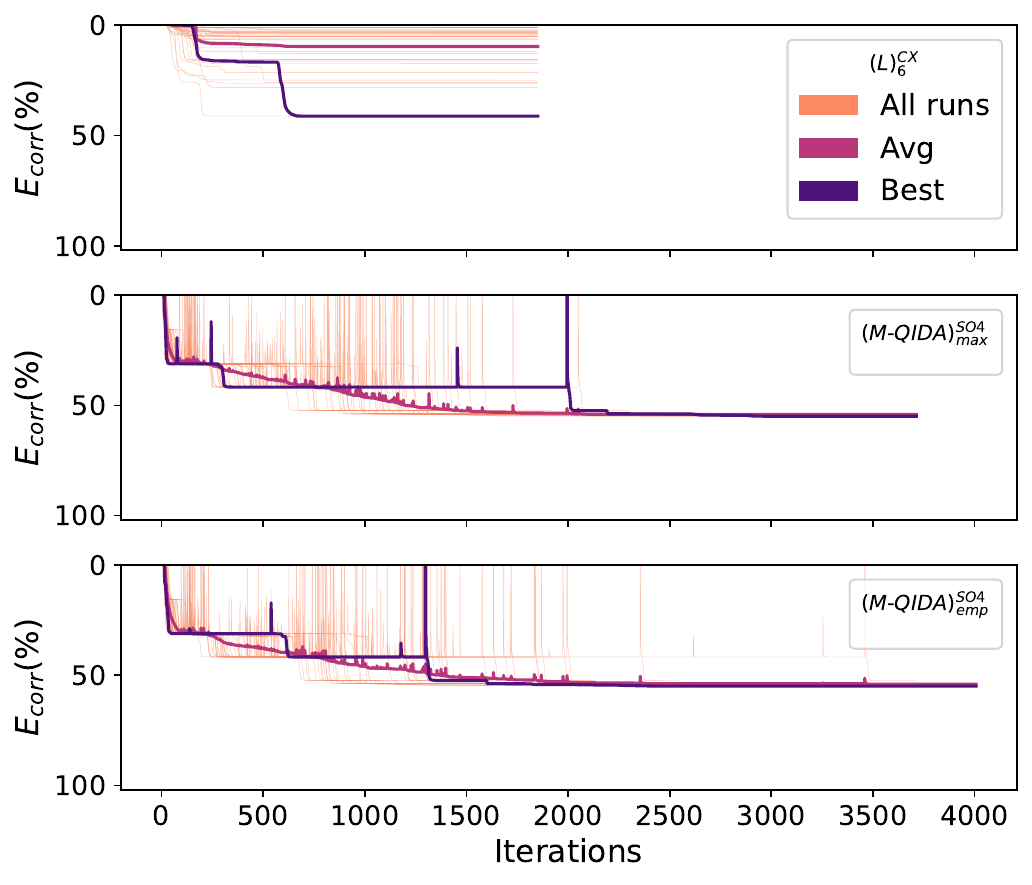}
    \caption{NH$_3$ INOs system convergence trajectories}
    \label{fig:nh3_ino_conv}
\end{figure}

Optimization trajectories for each of the CASCI/Active Region system of 50 VQE for HEA ladder-fashion circuit against Multi-QIDA circuits. 
\begin{figure}[!htb]
    \centering
    \includegraphics[width=7cm, height=5.5cm]{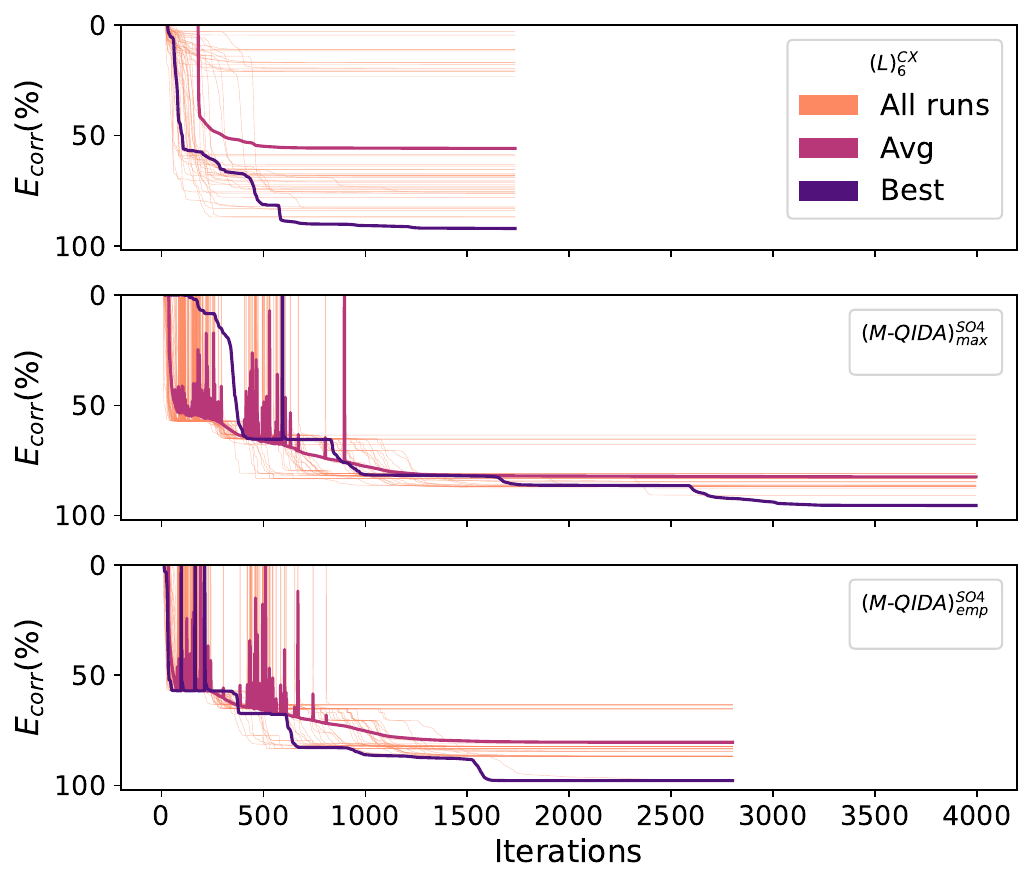}
    \caption{H$_2$O 6-31G CAS(4,4) system convergence trajectories.}
    \label{fig:h2o_cas_conv}
\end{figure}

\begin{figure}[!htb]
    \centering
    \includegraphics[width=7cm, height=5.5cm]{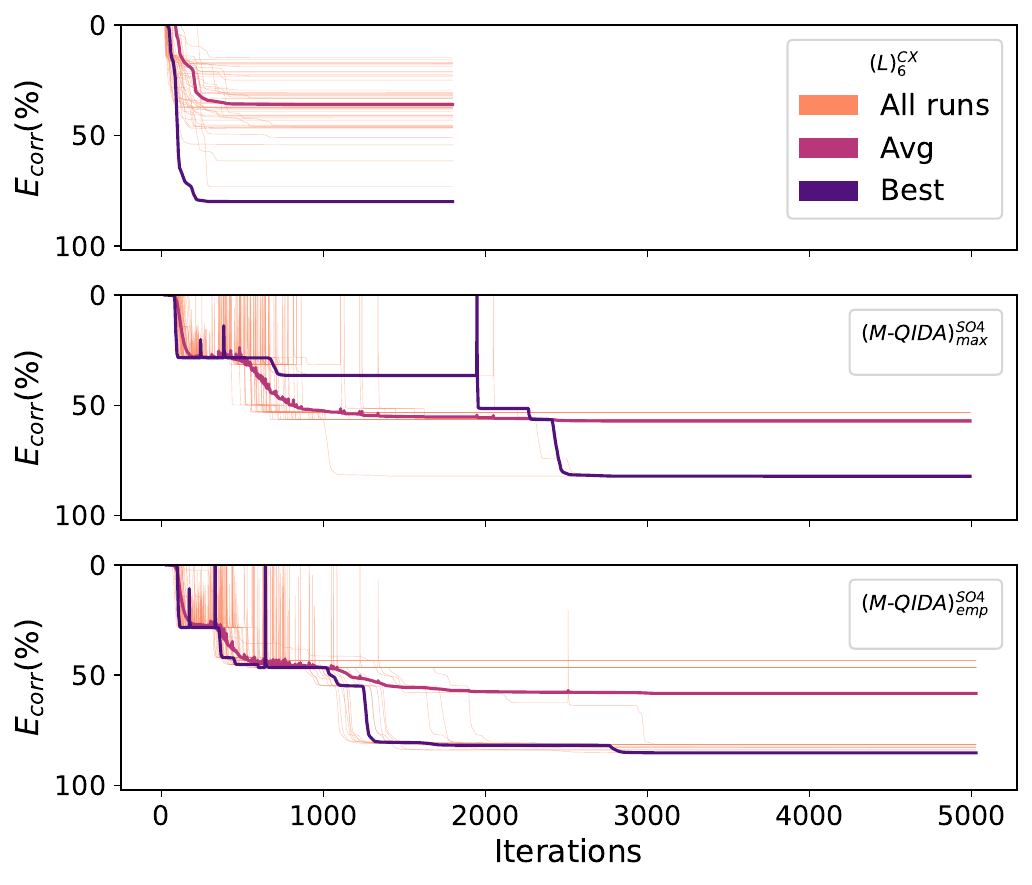}
    \caption{$N_2$ cc-PVDZ CAS(6,6) system convergence trajectories.}
    \label{fig:n2_cas_conv}
\end{figure}
\end{appendices}
\end{document}